\documentclass[iop,revtex4]{emulateapj}

\usepackage{ifpdf}
\usepackage{pdflscape}
\usepackage{hyperref}
\usepackage{doi}
\usepackage{threeparttable}
\usepackage{longtable}

\newcommand{\SETHI}{{SETH{\sc i}}}
\newcommand{\kms}{km s$^{-1}$}
\newcommand{\HI}{\ion{H}{1}}

\newcommand{\etal}{et~al.}

\newcommand{\unit}[1]{\ifmmode {\rm\ #1} \else {$\rm #1$} \fi}

\newcommand{\degrees}{\ifmmode{^{\circ}}\else{$^{\circ}$}\fi}
\newcommand{\degree}{\ifmmode{^{\circ}}\else{$^{\circ}$}\fi}
\newcommand{\ifpp}[1]{#1}	
\newcommand{\ifms}[1]{}	

\ifpp{
\newcommand{\spc}{\,}

\newcommand{\ads}[1]{ ADS:\href{http://adsabs.harvard.edu/abs/\detokenize{#1}}{\ttfamily \detokenize{#1}}\spc}
}

\slugcomment{Accepted by{\it The Astronomical Journal}}

\shorttitle{\SETHI\ Catalog of Interstellar \HI\ Shells}
\shortauthors{Sallmen et al.}

\begin{document}


\title{Interstellar \HI\ Shells Identified in the \SETHI\ Survey}


\author{Shauna M. Sallmen\altaffilmark{1}, Eric J. Korpela\altaffilmark{2},
Brooke Bellehumeur\altaffilmark{3}, Elizabeth M. Tennyson\altaffilmark{4}, 
Kurt Grunwald\altaffilmark{1}, Cheuk Man Lo\altaffilmark{5}}
\email{ssallmen@uwlax.edu}
\altaffiltext{1}{University of Wisconsin - La Crosse, La Crosse, WI 54601}
\altaffiltext{2}{Space Sciences Laboratory, University of California at Berkeley, Berkeley, CA, 94720}
\altaffiltext{3}{Milwaukee, WI}
\altaffiltext{4}{Materials Science and Engineering Department, University of Maryland, College Park, MD 20742}
\altaffiltext{5}{Hong Kong, China}


\begin{abstract}
Galactic \HI\ (neutral hydrogen) shells are central to our understanding
of the interstellar medium (ISM), which plays a key role in the 
development and evolution of galaxies, including our own. Several
models involving supernovae and stellar winds have contributed to our
broad understanding, but a complete, detailed picture remains elusive.
To extend existing Galactic shell catalogs, 
we visually examined the \SETHI\
(Search for Extraterrestrial \HI) database to identify shell-like
structures. This high-sensitivity 21-cm radio 
survey covering the Arecibo sky uniquely provides high-resolution
data on shells at a wide range of Galactic latitudes.
We present basic information (location, radial velocity, angular size,
shape) for 74 previously unidentified \HI\ shells. 
Due to limitations of coverage and data quality,
and the biases inherent in search techniques, our catalog is not a complete
sample of Galactic shells.
We discuss the
catalog completeness, and comment on the new shells' relationship with
known interstellar structure as warranted.  Unlike many previous 
catalogs, this sample is not biased towards expanding shells.  Where 
possible we also estimate the kinematic distances, physical sizes, 
expansion velocities, and energies of these shells. Overall,
they are relatively large and old, each the result of multiple
supernovae. Unlike previous surveys, we do not find that the shells
in our sample are preferentially aligned relative to the Galactic plane.

\end{abstract}




\section{Introduction}

The interstellar medium (ISM) plays a key role in the development and
evolution of galaxies, including our own. The effects of generations
of stars within the galactic ISM have produced a turbulent, multiphase
medium filled with complex interacting structures. Shells, bubble-like
features, ``chimneys" and ``worms" were first identified in neutral
hydrogen (\HI) maps by \citet{heil79, heil84}. These structures are
driven by stellar winds and supernova (SN) explosions.  These processes
are responsible for redistributing energy and material throughout our
galaxy, resulting in the formation of new generations of stars.

The physical state and evolution of these gas phases are likely explained
(at least in part) by the three-phase model of \citet{mckost77}, wherein
random supernovae result in a turbulent ISM of hot, low-density gas
surrounding warm and cold clouds.  In the Galactic fountain model of
\citet{shap76}, hot gas rises out of the Galactic plane, cools, then
falls back into the Galactic plane.  Superbubbles (caused by clusters of
supernovae) can break out of the Galactic plane providing a source of bouyant
hot gas to a galactic fountain.  The extent to which this affects the overall
structure and distribution of the gas is unclear.

Agreement remains elusive when it comes to details such as the filling
factor of the various phases, and porosity of the medium \citep[see
reviews by][]{cox05,ferr01}.  These depend on the number and energy
distribution of supernova events, and how they interact with the
surrounding medium. The role of magnetic fields in the interaction is
unclear, although some models of supernova evolution have incorporated
their effects \citep[e.g.][]{slavcox92}.  In the \citet{slavcox93}
picture, for example, the disrupting influence of supernovae is relatively
small. However, the energy inputs of shells are imperfectly understood.
The number and size of large shells in the outer galaxy cannot yet be
explained by the expected level of star formation in those regions,
despite consideration of numerous alternatives \citep[see][]{mcc-g02}.

Because \HI\ shells are central to our understanding of the ISM, it is
important to identify shells at all stages of evolution for further
study.  Early shell identification \citep{heil79, heil84} was based
on visual inspection of data, and so included non-expanding shells.
Later searches for shells \citep{mcc-g02, ehl05, djp07} 
commonly used expansion as one of the criteria for shell identification.
This had the advantage of discriminating against random superpositions
of filamentary gas, but the disadvantage of biasing the shell catalogs
against older, more evolved shells. The most recent searches 
\citep{EP2013,Suad2014} do include non-expanding structures, but
are based on relatively low-resolution data. Previous searches
carried out in high-resolution data are restricted to within
a few degrees of the Galactic plane \citep{mcc-g02, djp07}.

In this paper, we present shells found in a visual-identification
search of high-resolution data, in order to extend the Galactic census
of \HI\ shells.  The \SETHI\ (Search for Extraterrestrial \HI) dataset
is described in Section~\ref{sect:sethi}.  We describe the search
methodology and discuss the completeness of our search in Section~\ref{sect:method}.
We present our search results in Section~\ref{sect:results} and 
compare our findings to those of other surveys in Section~\ref{sect:comparecat}.  
In Section~\ref{sect:discussion} we discuss the physical properties of the 
shells and what our observations tell us about shells in our Galaxy.

\section{Description of \SETHI\ Survey}\label{sect:sethi}

When this work was performed, the \SETHI\ survey was the single-dish 
large-scale survey with the highest angular resolution. 
Although 
interferometric surveys have higher resolution, the typical sensitivity of 
the \SETHI\ survey exceeded that of the available interferometric
surveys. \SETHI\ also covered a larger range of Galactic latitudes than
interferometric surveys, which are typically limited to within a few degrees 
of the Galactic plane.  The Galactic Arecibo L-Band Feed Array \HI\ 
\citep[GALFA-\HI;][]{peek11}
Survey was not available when this project began, and until 
its \HI\ Data Release 2, expected in mid- to late-2016, 
still has less complete sky coverage than the \SETHI\ survey.

\SETHI\ was an outgrowth of the 
University of California, Berkeley (UCB) SETI program.  
Between 1999 and 2006 these searches used an uncooled 
receiver on the 1420-MHz flat-feed 
on Carriage House 1 at the
National Astronomy and Ionospheric Center's 305-meter radio telescope in
Arecibo, Puerto Rico.
This carriage house is opposite the zenith from the primary receivers in the
Gregorian dome, allowing observations covering most of the Arecibo sky
to be conducted while not interfering with other uses
of the telescope.  This resulted in two main modes of observation.  When the
primary feed was stationary or stowed the flat-feed beam 
scanned across the celestial sphere
at the sidereal rate.  If the primary observer's feed was tracking a position
on the celestial sphere, the Carriage House 1 beam scanned the sky at appoximately twice the 
sidereal rate.  At twice
the sidereal rate, the 6\arcmin\ half power beam width corresponds to a 12
second duration for a source to cross the beam.
Over the duration of this survey, a large majority of the sky visible to the
Arecibo telescope was covered.  \citep{korpela02,korpela04}.  

The time domain data for the sky survey were recorded as follows:  first,
a 30-MHz band from the receiver was converted to baseband using a pair
of mixers and low-pass filters.  The resulting complex signal was digitized, 
then filtered to 2.5MHz using a pair of 192 tap FIR filters in the SERENDIP IV 
instrument \citep{werthimer97}.
Single bit samples (one real and one imaginary bit per complex sample) 
were recorded on 35-GB DLT tapes.   
  These were shipped to Berkeley for the SETI@home program.  

The \SETHI\ survey analyzed these tapes to extract hydrogen spectra.
The 2.5-MHz time series data were converted to raw spectra using 2048 point 
FFTs ($\Delta\nu$=1220 Hz).  We then accumulated 6144 spectra into a single 
power spectrum with an integration time of 5.033 seconds.  
The resulting power spectrum was corrected for one-bit sampling
effects by applying the Van Vleck correction.  The spectrum, its start and
end coordinates, and the observation time were stored in a database.

Because no absolute power calibration was available in the receiver or recorder
subsystem we calibrated our observations using existing surveys.  We first
performed a polynomal fit to remove broadband background
variations, then implemented a system temperature calibration   
by performing a linear fit of the \SETHI\ spectra to spectra from the
Leiden-Dwingeloo survey \citep[LDS;][]{lds97}.  While this method has
the drawback of reducing our sensitivity to changes in {\bf total} \HI\ column density
on scales smaller than the LDS beam size (36\arcmin), changes in
the spectral velocity profile are well preserved on scales near the \SETHI\ beam size.
The more effective calibration developed for the GALFA-\HI\ Survey \citep{peek11},
requires some observing techniques that are inapplicable to \SETHI.

The spectral fitting resulted in an estimate of the system temperature (including any
background continuum components).  Our system temperatures fell between
60 and 170 K approximately 65\% of the time.  Excursions outside of this range 
due to receiver problems or excessive noise environments 
resulted in unusable data which we excluded from further processing. 

The \SETHI\ data were accumulated into 256$\times$256$\times$1591-pixel data
cubes (RA, Dec, V$_{LSR}$) of dimension 7.68\degrees$\times$7.68\degrees$\times$410 \kms.  
Pixel dimensions in these cubes are 0.03\degrees$\times$0.03\degrees$\times$0.26 \kms.

\tabletypesize{\scriptsize}
\begin{table}[tbp]
\begin{center}
\caption{Comparison of the parameters of the LDS and \SETHI\ surveys\label{comp}}
\begin{tabular}{lcc}
\hline
\multicolumn{1}{c}{Parameter} & Leiden/Dwingeloo & \SETHI \\
\hline\hline
Angular Resolution (HPBW) & 0.6\degrees & 0.1\degrees \\
Spectral Resolution & 1.0 \kms & 0.25 \kms \\
Spectral Range & 1000 \kms & 520 \kms \\
Sensitivity & 0.07 K & 0.25 K \\
Sky Coverage &  $\delta > -30\degrees$ & $7.2\degrees < \delta < 29.7\degrees$\\
\hline
\end{tabular}
\end{center}
\end{table}

Table~\ref{comp}\ shows a comparison of the parameters of the LDS and
\SETHI\ surveys.  The Leiden/Argentine/Bonn \HI\ \citep[LAB;][]{kal05}
survey's sensitivity and resolution (both angular and spectral) are not
significantly different from those of the LDS.  While having superior
spatial and spectral resolution, the \SETHI\ survey's spectral range,
sensitivity and sky coverage are inferior to the LAB/LDS survey.  If both
surveys were at the LAB/LDS angular and velocity resolution scales,
\SETHI's sensitivity is potentially 33 times better than quoted.
The SETHI survey is prone to artifacts matching the grid spacing
of the LDS spectra, because of the limits of single-bit data and the
cross-survey calibration mechanism.  Nonetheless it is useful for finding
faint features on scales smaller than the LAB/LDS survey resolution.

\section{Method}\label{sect:method}


Because the \SETHI\ survey covers a constant-declination strip of
the sky, we performed our search in equatorial coordinates.  For the
purposes of this study, we merged the standard \SETHI\ cubes into larger
overlapping cubes.  Each was 22.5\arcdeg\ on a side, with a pixel size
of 0.03\arcdeg\ and a velocity resolution of 1.55 \kms.  These cover
the full range of RA, and are centered at a declination of 18.5\arcdeg.
The velocity range was restricted to exclude distinctly extragalactic gas,
extending from -119.02 \kms\ to 289.35 \kms.

\ \par
\subsection{Why Visual Identification?}

With the increased velocity and angular resolution of current surveys,
searching by eye is a daunting task.  Nonetheless, purely automated 
searches are extremely difficult, as they work best for closed, regular,
expanding structures. 
For example, the method of \citet{mash02} and \citet{mash99} 
is based on the results of hydrodynamic models, and is problematic due
to the non-uniformity of the ISM (e.g. fragmented or non-spherical shells,
variations in background or foreground emission). 

At the start of our search process, the most complete automated search 
for shells to date was the
catalog of \citet{ehl05} based on the LDS, which covered 79\% of the sky at an
angular resolution of 0.5\arcdeg.  The catalog contained 
only closed structures with signs of expansion based on their central spectra,
and selected against low-contrast shells in a rapidly changing background.
Its completeness 
was best for younger shells (smaller, not blown out, not very distorted),
but it was very incomplete for fragmented older shells 
with minimal expansion. For such large, irregular, non-expanding 
and possibly open structures, visual identification was the preferred technique
\citep{ehl05}. This search was recently updated to eliminate the
requirement that shells show evidence of expansion, and extended to the 
whole sky using the LAB survey \citep{EP2013}. Their algorithm still requires
closed structures, meaning fragmented older shells are likely to be missed.

The higher angular resolution of the \SETHI\ data means the incomplete
nature of shell walls is more evident, making a visual search
very appropriate.  In addition, automated searches are much more
computationally intensive for high-resolution data, and are frequently
limited to younger expanding shells. The completely automated search of
\citet{djp07} used neural networks to identify small expanding shells
in high-resolution data of a small (48\arcdeg $\times$ 9\arcdeg)
section of sky.  Although non-uniformity of the ISM and possible
extreme variations in shell shape make visual identification of shells
subjective, the same issues also make it difficult to specify appropriate
criteria for automatic searches \citep{djp07}.  Recently \citet{Suad2014}
utilized a visual search to train an automated search algorithm focusing
on supershells in lower-resolution data, acknowledging that "the eye is
an incredibly powerful instrument, especially when images are irregular".

Due to the complex structures visible in our high-resolution data, and
since every search technique requires visual inspection at some stage,
we chose to take advantage of the human visual system.
The procedure we followed does not require that the shell be expanding,
although shells are deemed to be of higher ``quality" if signs of expansion
are present. Similarly, it allows for the identification of partial or
fragmented shells.  Care must be taken, however, to maximize consistency
in shell identification and classification.

\subsection{Description of Search Process \& Criteria}\label{subsec:process}

The merged \SETHI\ data cubes were viewed using the $<$kvis$>$ program
of the Karma Toolkit \citep{gooch96}.  To minimize errors and missed
features, a minimum of two undergraduate students searched each cube
using scalings chosen to highlight features in regions of both strong
and weak emission.  This resulted in a tentative list of 80 shells,
28 of which we eliminated as previously cataloged \citep{ehl05, eph01,
heil84, hu81, heil79}.  Fifteen (15) additional features were removed
after further inspection by Korpela and Sallmen. We then searched all
data cubes for missed shells, adding 28 convincing unknown
shells to our list. Data cubes including the Galactic plane were given
extra attention.  During our exploration our catalog's completeness
(Section \ref{sect:completeness}), we found 9 additional shells and
added them to our catalog.

For all 74 shells in our final
catalog we identified the velocity at which the shell is most
evident (V$_{\rm ref}$) and estimated the position of its center (RA,
Dec), angular extent ($\Delta$RA, $\Delta$Dec), and the range of velocities
over which it appears ``shell-like" ($V_1$ and $V_2$).  Typical errors in
central and edge position estimates are about 0.25$\arcdeg$ for small shells
(less than 5$\arcdeg$ across), 0.5$\arcdeg$ for large shells (5 to 10$\arcdeg$
across), and up to 1$\arcdeg$ for very large shells (greater than 10$\arcdeg$
across).  For a few shells with one edge near the 
declination limits of our data, we used the LDS data to estimate $\Delta$Dec.
Note that V$_{\rm ref}$ may not be 
the central velocity of an expanding shell, but was defined to
be valid even for old, non-expanding shells. 

To determine mean angular diameters ($\Delta\theta$) for each shell, we 
identified the four locations at the ends of the long and short axes, then
averaged the axis length estimates ($\Delta\theta_1$ and $\Delta\theta_2$). 
We define a shape parameter 
$S = 1 - (\Delta\theta_1 - \Delta\theta_2)/2\Delta\theta$ to desribe the
elongation of the shell. Note that 
$S$ = 0 for a line, and $S =$ 1 for a perfect circle.  However irregular 
shells often lack well-defined short and long axes, making these
measurements subjective on shells of that type, particularly for 
triangular or complex shapes.
To quantify likely uncertainties, $\Delta\theta$ and $S$ estimates made by 
multiple (two to five) people were compared for nearly all shells.
Approximately 80\% have differences in mean angular diameter of $<$ 10\%; 
the same was true of the shape parameter for $\sim$85\% of shells.
Reasonable differences of opinion larger than this (but $\lesssim$ 20\%) 
occur mostly for shells that are relatively small, have very thick or irregular 
walls, or more than one of these characteristics. 

\begin{figure}[tbp]	
\ifpdf
\includegraphics[width=\columnwidth,keepaspectratio=true]{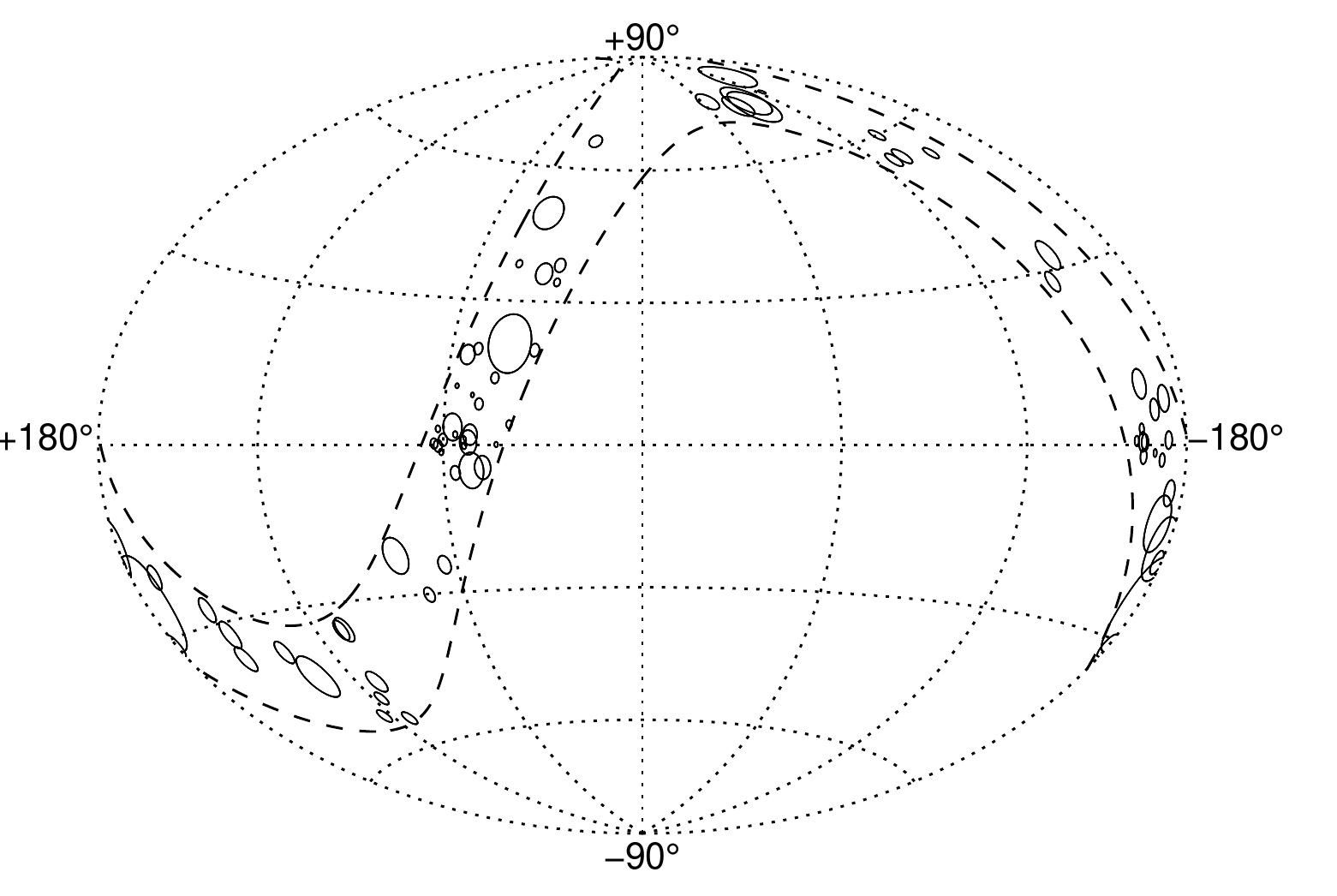} 
\else
\includegraphics[width=\columnwidth,keepaspectratio=true]{f1.eps} 
\fi
\caption{Sky plot of 74 shells newly identified in the \SETHI\
dataset, in Galactic coordinates. The dashed line shows the limits of 
the \SETHI\ data. Circles reflecting shell locations and mean angular
diameters in equatorial coordinates were projected onto this representation.
\label{fig:skyplot}}
\end{figure}

Figure \ref{fig:skyplot} shows the locations and sizes of these 74 shells
on an Aitoff projection of Galactic coordinates. For each shell, we
projected a circle with diameter equal to its mean angular diameter 
$\Delta \theta$ in equatorial coordinates.
Our search identified a number of relatively small shells at a wide range
of Galactic latitudes, due to the unique characteristics of the \SETHI\ survey.
Near the Galactic plane, many new small shells were identified, but 
few large ones remained unknown in this well-studied part of our Galaxy.

To determine the ``quality" of each shell, we estimated the following quantities.
\begin{enumerate}
\item The fraction ($f_{\rm closed}$) of the shell which is ``closed" at V$_{\rm ref}$,
taking into account the weight/strength of the walls. $f_{\rm closed} = 1$
for a shell which is 100\% clearly and evenly closed. 
\item The shape parameter $S$ defined above.  More regular shells are more believable.
\item The consistency ($C$) of the feature's shape and location 
across the range of velocities (ignoring size changes).
$C=1$ implies both consistent shape and location. 
High scores indicate persistence in velocity, making the structure less likely 
to be a random overlay of disconnected features. 
\item Fraction ($f_v$) of the velocity range over which the shell completeness 
remains at the value $f_{\rm closed}$. 
Higher values result in more convincing shells. 
If one wall is visible at velocities where the others are not, 
the shell might be strongly sheared, or the structure might not 
be a shell.
\item Whether the shell appeared smaller in maps towards the velocity extremes, 
and largest near the central velocity. Half of the consistency score 
($C_{\theta}$) was allotted to each end of the velocity range. 
$C_{\theta}=$1 is a clear signature of an expanding shell, while 
$C_{\theta}=$0 if the shell size remains unchanged 
(or does not do what is expected).  Intermediate values were based on a 
qualitative perception of how convincing the size changes were.
\item Whether any sign of expanding front and rear walls of the shell were 
visible in the maps at higher and lower velocities. Half of the wall detection ($W$) 
score was allotted to each cap.  $W=1$ indicates clear and unambiguous identification 
of both front (approaching) and rear (receding) walls, while
$W=0$ indicates no evidence of either wall, and no possibility of confusion
limiting our detection ability.  Intermediate values were based on 
qualitative perceptions of confusion levels and wall identifications.
\item Did we observe any sign of expansion in position-velocity (PV) space, 
i.e. in the Vel-RA and Dec-Vel maps?  For each of these, we assigned 
a score from 0 to 1 based on the credibility of velocity splitting, and
its consistency with previously determined size indicators. ($PV$: 0.3 for
any splitting, 0.5 for matching range in position and velocity, 
0.2 for maximum velocity in shell center). Scores for the Vel-RA and 
Dec-Vel indicators were averaged to produce the final value.
\end{enumerate}

We combined the first four parameters ($f_{\rm closed}$, $S$, $C$, and $f_v$) 
into an overall quality estimate ($Q$), and the last three parameters 
($C_{\theta}$, $W$, and $PV$) into an expansion quality estimate ($Q_{\rm exp}$),
as the former do not depend on shell expansion.  To facilitate their 
comparison, both $Q$ and $Q_{\rm exp}$ were scaled to a maximum value of 10. 
Scatter plots and correlation estimates of each parameter against both 
quality scores confirmed that this division was appropriate. Other
measured quantities (such as mean angular diameter or the velocity range over
which the shell appeared shell-like) did not correlate with either quality
score.  These quality estimates do not necessarily reflect how
convincing a shell looks at its reference velocity, because low scores for its
shape and/or wall strength at the reference velocity may be offset by high 
scores in other contributing factors.

After completing the catalog and much of the analysis below, new \HI\ shell 
lists were published by \citet{EP2013} and \citet{Suad2014}. We have examined
the overlap between our new list and these lists, but did not remove
any additional shells from our catalog. We discuss specific cases of 
commonality in Section \ref{subsec:comments} and the overall comparison in 
Section \ref{sect:comparecat}.

\subsection{Exploring Catalog Completeness}{\label{sect:completeness}

The process of visual search and identification is necessarily subjective,
as \HI\ sensitivity is not even across the survey, and 
shell detection depends on image display settings and individual
perceptions. Although we made a point of having at least two
individuals search each data cube at a variety of scalings, interstellar
features could have been missed.
The completeness of our search was explored by choosing four data cubes
and re-examining them thoroughly. Two of these cubes included the Galactic
plane, while the others were at high Galactic latitudes. For each, Korpela
\& Sallmen strived to identify all potentially shell-like features,
and evaluated whether we would include them in the catalog using
the categories Yes, No, and Possibly. We then compared this feature list
with existing shell catalogs, as well as new \SETHI\ shells 
identified during our original search.

During this second-look process we identified 32 features as worthy of
catalog inclusion. Of these, 11 were already in our catalog, and 12 others had
previously been published. However, we added 9 new shells to our
catalog; 6 in the Galactic plane and 3 at high latitude. Similar extra
attention to the remainder of the \SETHI\ data would likely result in
additional entries, but the number is difficult to quantify. We also 
identified 33 features as possibly worthy of inclusion, 22 of which 
we either failed to identify or rejected during our initial search. 
These results
illustrate the subjective nature of our criteria for catalog inclusion
of borderline shell-like features, but we did not add these to
the catalog.

Prior to this second look, we had previously cataloged 
31 shells with centers in these four SETHI
cubes.  However 3 were near the cube edges, so more easily recognized
in adjacent overlapping data cubes.  During our second-look process, we
re-identified 22 of the remaining 28 shells, and categorized 19 of them
as Yes or Possibly worthy of catalog inclusion. Some of the remaining
catalog shells are complex features that overlapped or merged with other
interstellar features when stepping through the velocity slices.
For others we noted poor contrast or
wall definition, suggesting they were on the borderline in our previous
evaluations.  Nonetheless, some shells are missed each time a cube is
searched, validating our use of multiple individuals searching the data.

The \SETHI\ catalog is subject to several selection effects. Firstly,
our shell detection rate depends on shell size. Shells bigger than
10\degree\ to 15\degree\ are difficult to identify in our 22.5\degree\ data
cubes, especially if they cross cube boundaries.  Although we explored
overlapping data cubes to mitigate this issue, shells extending beyond the
data's Declination limits are not detectable. This especially limits the
detection of large shells. Despite the data resolution, shells smaller
than about 3\degree\ become more difficult to identify due to the large
size of the search region. For such small shells, only those with sharp
edges and high contrast are likely to be spotted. This effect makes
shells smaller than about 1\degree\ very difficult to detect through
our search method.

Our sensitivity to interstellar features also varies slightly across the 
dataset, partly due to the involvement of multiple students
and the long time frame of the search, but also due to variations
in the appearance of Galactic gas in different directions in the regions
available to the Arecibo telescope.  Because we
expect shells to be more numerous in the Galactic plane, we ensured that
data covering those regions received extra attention. These shells are
extremely useful as potential targets for future study because they
have low Galactic latitudes, so are more amenable to kinematic distance
estimation.  Despite the additional consideration, the completeness here
may be somewhat less than other regions, due to the increased complexity
of interstellar gas.  Finally, the velocity range of the data
does not fully cover Galactic gas; our data exclude some very distant
spiral arms which may have different properties from more local ones.

\section{Results}\label{sect:results}

Table  \ref{table:shell_list} presents the \SETHI\ shell catalog.
Columns 1-4 contain the Shell identifier, equatorial coordinates (RA,
Dec), and reference velocity at which it is most 
clearly identifiable ($V_{\rm ref}$). Columns 5-8 describe the shell's
spatial and velocity extent ($\Delta$RA, $\Delta$Dec,
$V_1$, $V_2$). Columns 9-10 contain the mean angular diameter
($\Delta\theta$) and shape parameter ($S$). Columns 11 and 12 contain numerical
estimates of the overall shell quality ($Q$ and $Q_{\rm exp}$). Flags in
column 13 point the reader to additional comments (see Section 
\ref{subsec:comments}).

All search results in Tables \ref{table:shell_list} through \ref{table:PA} 
are available at \url{http://setiathome.ssl.berkeley.edu/\~shauna/SETHI\_Shells.html}, along with links to images and summary information for each shell.
The information provided assists the \SETHI\ data and catalog
user in identifying shells within the equatorial-coordinate maps.
The \SETHI\ data cubes may also be retrieved here.


The HI maps for 6 shells selected from our catalog are shown in Figures 
\ref{fig:shellimages-1} and \ref{fig:shellimages-2}.  Each is described 
briefly below.  

\begin{figure}[tbp]	
\begin{center}
\ifpdf
\includegraphics[width=0.99\columnwidth,keepaspectratio=true]{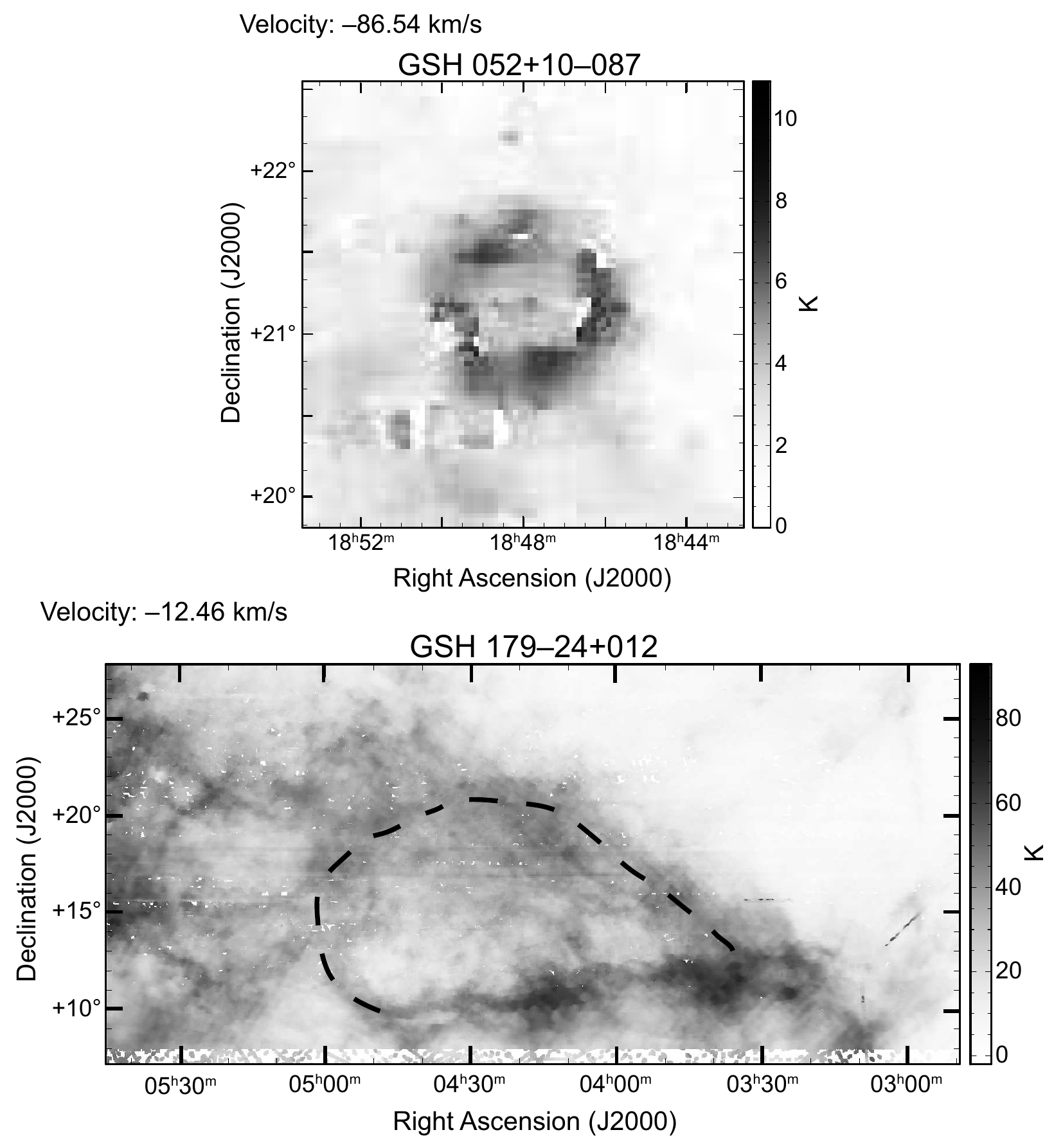}
\else
\includegraphics[width=0.99\columnwidth,keepaspectratio=true]{f2.eps}
\fi
\end{center}
\caption{Neutral hydrogen maps for shells GSH 052+10-087 (top), and 
GSH 179-24+012 (bottom). Each shell is shown at its reference velocity. 
Darker shadings indicate regions of greater \HI\ emission. The dashed line
for GSH 179-24+012 indicates the extent of the shell in places where
confusion might otherwise occur.
\label{fig:shellimages-1}}
\end{figure}

GSH 052+10-087 is the smallest shell in our catalog, with a mean angular 
diameter of $\Delta\theta = 1.0\arcdeg$.  It is extremely round ($S=0.95$),
high in quality ($Q = 8.7$; $Q_{\rm exp} = 6$), and 
lies in high-velocity gas ($V_{\rm ref} = -87$ \kms). 

GSH 179-24+012 is the largest shell in our catalog 
($\Delta\theta = 15.9\arcdeg$). Its teardrop shape ($S = 0.72$) 
extends from 01$^h$23$^m$ to 05$^h$00$^m$ in RA and from 9.75$\arcdeg$ to 
21.5$\arcdeg$ in declination.  

GSH 188+07-079 also lies in high-velocity gas, and has an
irregular peanut shape. Its mean angular diameter 
($\Delta\theta = 4.5\arcdeg$) is just above the average for our catalog.
This moderate-quality ($Q = 4.6$; $Q_{\rm exp} = 2.7$) 
shell is clearly non-spherical, giving it a low 
shape parameter ($S = 0.51$).

GSH 052-05+023 is an example of a shell that extends outward from the 
Galactic plane gas of a spiral arm.  Based on our high-resolution data, 
we identified the extension at the northeast end as
a separate shell (GSH 056-06+033).  Both quality parameters are high
($Q = 7.5$, $Q_{\rm exp} = 6.3$), as expected for a distinct
part of a previously known expanding shell (see discussion in the next
section).  It is also one of the few shells for which we can estimate both
kinematic distance and expansion velocity (see Sections \ref{sect:kind}
and \ref{sect:vexp}).

GSH 052+02-071 is a low-contrast shell straddling the Galactic plane. 
Like many shells in our catalog, it shows little sign of expansion 
($Q_{\rm exp} = 1.8$),
but has a high quality ($Q$ = 7.3) based on its other characteristics.

GSH 225+55-005 ($\Delta\theta$ = 3.4$\arcdeg$)
is embedded in a region of complex gas. Because it is quite irregular in 
shape, its high shape parameter ($S$ = 0.92) 
is misleading. It scores moderately well on visual
parameters ($Q$ = 6.8), but shows little evidence of expansion 
($Q_{\rm exp}$ = 0.8).  At $b = +55\arcdeg$, it lies well out
of the Galactic plane, unless it is within local gas.

\begin{figure*}[tbp]	
\begin{minipage}{\textwidth}
\begin{center}
\ifpdf
\includegraphics[width=\textwidth,keepaspectratio=true]{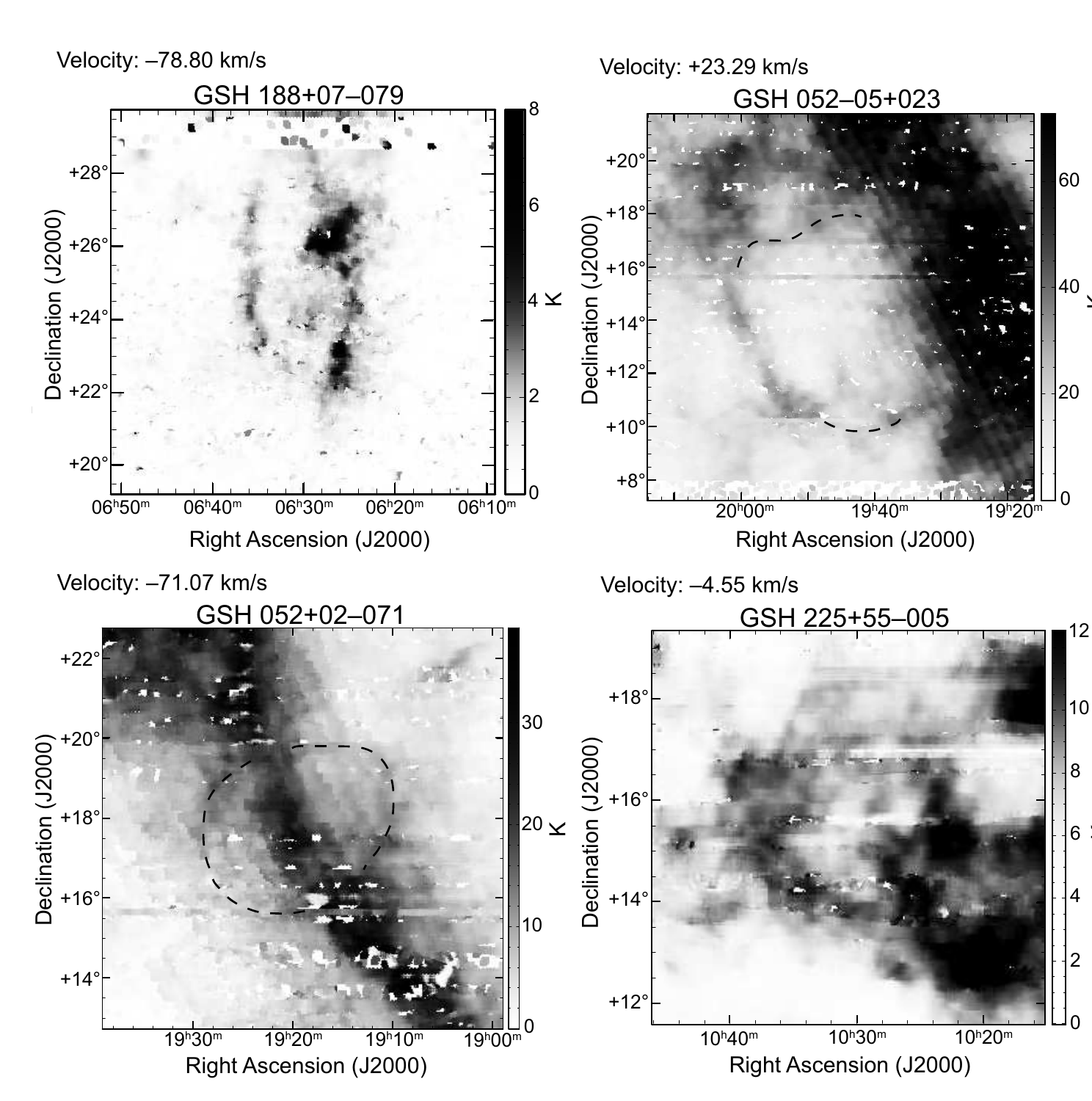}
\else
\includegraphics[width=\textwidth,keepaspectratio=true]{f3.eps}
\fi
\caption{Neutral hydrogen maps for shells GSH 188+07-079 (top left), 
GSH 052-05+023 (top right), GSH 052+02-071 (bottom left), 
and GSH 225+55-005 (bottom right).  Same as previous figure.
\label{fig:shellimages-2}}
\end{center}
\end{minipage}
\end{figure*}

\ \par
\subsection{Comments on Shells in Table \ref{table:shell_list}}\label{subsec:comments}

For shells identified in Table \ref{table:shell_list}'s Comment column,
noteworthy aspects are described below. This includes
cases where our data conflicted with published shell 
identifications.
\begin{itemize}
\item {\bf GSH 042+21+019} is a large structure related to the previously 
identified GSH 043+22+019, however \citet{ehl05} underestimated the extent 
of the shell. It is also not clear that their smaller GSH 037+19+016 is a 
separate entity.
\item Based on how it appears across its velocity range, {\bf GSH 048-05+045} 
is a large structure that contains GSH 047-03+040 
\citep[][small elongated feature at the southwest corner of our shell]{ehl05}, 
as well as Heiles Supershell GSH 048-04+049 
\citep[][east portion of our shell]{heil79}.  
At lower velocities, {\bf GSH 048-05+045} 
blows out into a larger (unseen) bubble. 
\item Lots of confusion and data artifacts are present in the vicinity of 
{\bf GSH 049+08+026}. Rather than being a shell, the feature might be composed 
of overlapping clouds.
\item {\bf GSH 052-05+023} and {\bf GSH 056-06+033} may be related to one 
another, and both may be part of GSH 049-05+020 \citep{ehl05} /
GSH 050.0-05.0+019.6 \citep{EP2013}, which may 
be Heiles shell GS052-05+25 \citep{heil79}.  However our high-resolution 
data suggest that these are two distinct features. 
\item {\bf GSH 056+02-074} may be one piece of a larger shell. 
There might be a smaller adjacent shell or an additional piece 
of the same shell at similar velocities to the northeast, but we could not 
determine definitively due to localized data quality issues.
\item \citet{ehl05} identified a feature inside of 
{\bf GSH 057+04+005} as their GSH 057+03+003. Our larger shell is round, but of 
low contrast so difficult to discern. 
\item Based on our data, {\bf GSH 062+00+045} contains
GSH 061.5-00.5+046.4 \citep{EP2013}, and also additional regions to 
the northwest.
\item The previously identified [EPH2001]62.1+0.2-18 appears to be part 
of our larger shell 
{\bf GSH 063+00-022}. The field of view of \citet{eph01} was only 
$4\arcdeg \times 4 \arcdeg$, and cut off about half of this shell.
\item {\bf GSH 064-24+011} is the rightmost of two \HI\ holes 
at similar velocities, possibly with breakthrough between them. 
\citet{ehl05} identified the combined features as a single shell,
although the  other half isn't very closed or shell-like.
Our feature is GSH 064.0-24.5+011.3 in \citet{EP2013}.
\item For {\bf GSH 134-43-062}, we chose $V_{\rm ref}$ = -61.79 \kms\ 
to maximize the feature's contrast. 
The north-south extent of the shell was estimated at -50.96 \kms, where 
the contrast is much lower but its upper and lower boundaries are 
identifiable.
\item {\bf GSH 155-32+005} is GSH 154.5-32.5+006.2 
\citep{EP2013}.
\item {\bf GSH 156-37-003} is the northwest portion of the feature
GSH 160.0-38.0-002.1 identified by \citet{EP2013}. A substantial
\HI\ wall across their feature delineates the southern edge of our shell.
\item {\bf GSH 157-27-045} contains GSH 158-27-039 of \citet{Suad2014}, 
as it is larger and more elongated than that structure.
\item {\bf GSH 180-31+020} is a relatively round shell with a dense diagonal 
stripe crossing it at the reference velocity. 
\item {\bf GSH 192+06-017} was first reported in \citet{korpela04} and
subsequently detected by \citet{EP2013}.
\item Although we chose to catalog them separately, {\bf GSH 197-02+034} 
and {\bf GSH 198+01+034} might be related to one another.
\item {\bf GSH 261+74-025} shares boundaries with the shell 
GSH 255+74-028 identified by \citet{ehl05}, but our data clearly show a 
larger shell.
\end{itemize}

During our search we removed from
further consideration shells previously discovered by others.
As noted above, we retained features when our data were clearly
in conflict with previous shell boundaries. In a few cases, we found
the divisions to be subsective so did not include the shells in our catalog, 
but describe the possible differences below.
\begin{itemize}
\item GSH 248+69-013 and GSH 250+68-005 were previously cataloged separately
by \citet{ehl05}, while our impression is that this is a single 
low-quality shell that morphs somewhat in shape, location and size at 
various velocities. 
\item GSH 046+09+010 \citep{ehl05} has two distinct lobes 
which may or may not be related. We would have considered cataloguing this as 
a single shell.
\end{itemize}

\section{Comparison with other surveys}\label{sect:comparecat}

Since features identified as previously known shells were not always
recorded by students during the initial search, we used the results of our
second-look process (described in Section \ref{sect:completeness})
to investigate how often we detected shells listed in catalogs
published prior to spring 2013. Of the 62 known shells with centers
in these four regions, we identified 17 during this examination of
the data. Approximately half of the other 45 were either too big to be
easily visible in our 22.5\degree\ cubes, or went off the edges of the
cube. We examined our data for the remainder ($\sim 20$), and attribute
their non-identifications to one or more of the following reasons: (1)
Poor contrast in our high-resolution data, particularly for shells from
\citet{ehl05}, as that search used $\Delta T$ while the \SETHI\ search
utilized $\Delta T / T$. (2) Poor shell closure and poorly defined walls
in our higher-resolution data, when compared with their appearance in the
lower-resolution data. (3) Shells were located
in complex regions of high confusion, or in regions where the \SETHI\
survey had poor data quality. (4) We rejected the feature because their
shape and/or size changed substantially with velocity, or they showed
poor persistence with velocity. For one or two shells, these factors
were so extreme we couldn't identify the published feature in our data
at all. Given the differences in our dataset and search criteria compared
with previous work, these results are not surprising.

After our catalog and the above analysis was complete, two additional
shell catalogs became available: \citet{EP2013} and \citet{Suad2014}.
We subsequently investigated potential overlap between these and our \SETHI\
catalog. Of the 74 shells in our catalog, only three were also identified
by one of these other searches.  Three additional \SETHI\ shells are clearly 
sub-sections of larger features in the other catalogs, and in
two cases the other searches identified features clearly contained
within a larger \SETHI\ catalog entry. For all of these, the \SETHI\
data reveal clearer shell boundaries for our classification choices.
Finally, five more \SETHI\ shells may be related to
entries in the other catalogs, although the connection is not 
clear given the \SETHI\ data.

Our catalog comparison findings echo those reported by \citet{EP2013}
and \citet{Suad2014}: altering the search method and/or selection criteria
strongly affects shell identification. They found this true even
for searches using data of similar angular resolution; our catalog is 
based on data of much higher resolution, so substantially different
results are to be expected. Our search is unique in exploiting
the \SETHI\ survey's high angular resolution over a large range
of Galactic latitudes.

We also explored the robustness of our shell identifications
in higher-resolution data using the VLA Galactic Plane Survey
\citep[VGPS;][]{Stiletal06}, which has lower sensitivity than the \SETHI\
survey.  Seven of our catalog shells are small enough ($\Delta\theta < 3\arcdeg$) 
and located within the VGPS data ($l < 65\arcdeg$; $|b| < 2\arcdeg$). 
For four of these, the VGPS data are
consistent with our analysis, but reveal more detailed shell
structure. Two 
are visible but not as obvious in the VGPS data, because 
the higher resolution accentuates wall fragmentation
and the presence of faint wispy material within the shell. In one case
(GSH 054+01+031), the high-resolution VGPS data suggest a very faint,
thin wall at $b < 0$ that is not obvious in our \SETHI\ data. The feature
we identified as the low-$b$ wall may be a cloud inside the shell.
To maintain consistency in our catalog, we have not revised our
measurements of this shell.

\section{Discussion}\label{sect:discussion}

\subsection{Physical Properties of Shells}

\subsubsection{Statistics of Shell Properties}

Figure \ref{fig:Qhist} shows the distribution of Quality Estimates $Q$
(solid) and $Q_{\rm exp}$ (dotted) for the 74 shells in our catalog. 
Most shells have relatively high overall quality estimates $Q$, reflecting 
the fact that we did not measure shells we found less convincing. Our 
catalog is therefore less complete at the lower end of the range of $Q$ values. 
However, most shells have low expansion quality estimates, validating
that our approach avoids a significant bias against older slowly 
expanding shells. 

\begin{figure}[tbp]	
\ifpdf
\includegraphics[width=\columnwidth,keepaspectratio=true]{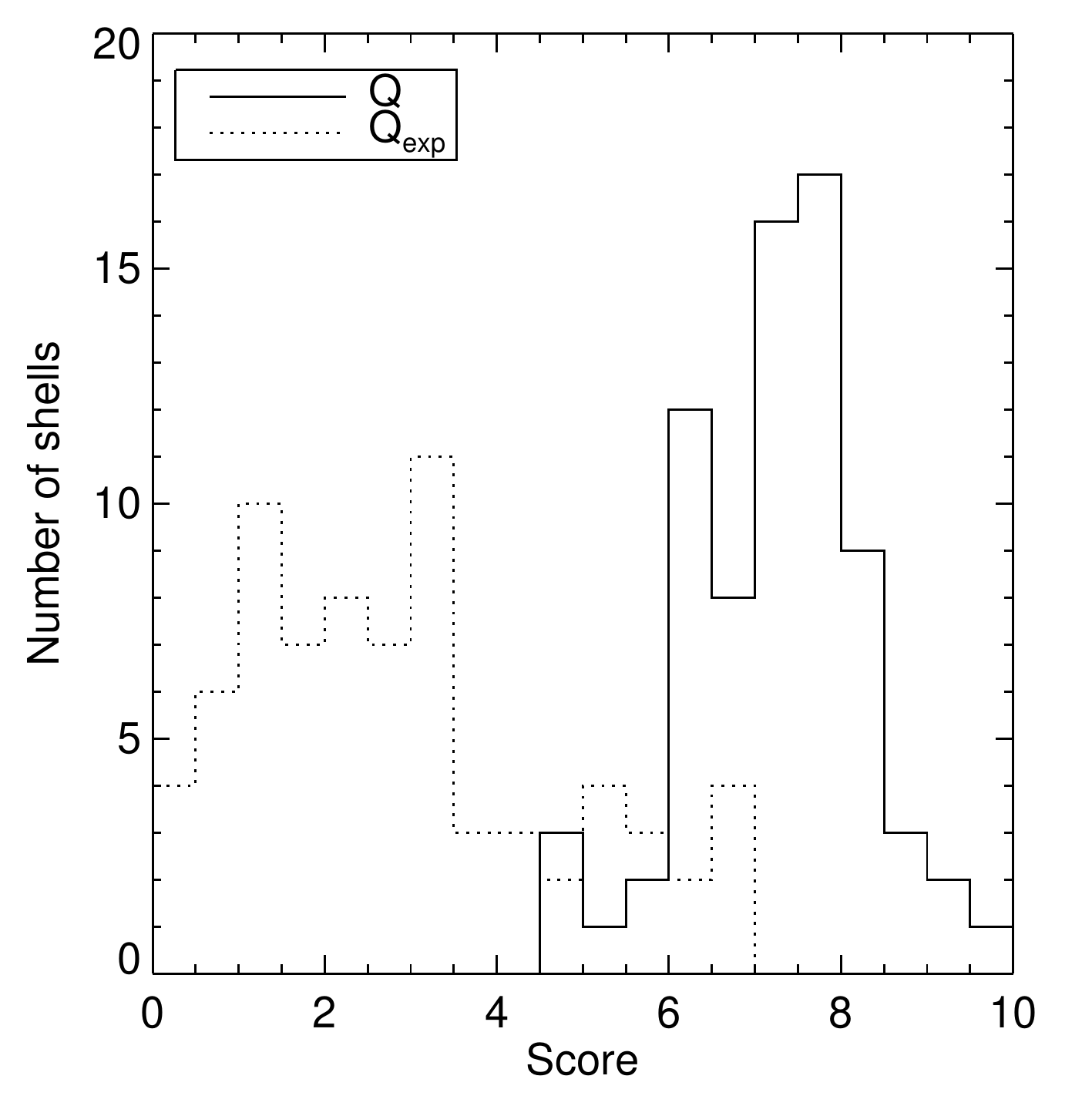}
\else
\includegraphics[width=\columnwidth,keepaspectratio=true]{f4.eps}
\fi
\caption{Distribution of Quality Estimates for the 74 shells in our catalog.
The solid line shows the histogram of overall quality estimates $Q$, which are
independent of evidence for shell expansion, while the
dotted line shows the histogram of expansion quality extimates $Q_{\rm exp}$.
\label{fig:Qhist}}
\end{figure}

Figures \ref{fig:MADhist} and \ref{fig:Shapehist} display the distribution of
mean angular diameters ($\Delta\theta$) and shape parameters ($S$) for
our catalog. The average mean angular diameter is 4.1$\arcdeg$, with a 
median value of 3.3$\arcdeg$. Although the distribution
extends to fairly large shells, most newly discovered shells are
relatively small. Most larger shells 
were discovered in previous lower-resolution surveys. The
lack of shells smaller than 1$\arcdeg$ is due to limitations in our survey
resolution, calibration, and search methodology.

\begin{figure}[tbp]	
\ifpdf
\includegraphics[width=\columnwidth,keepaspectratio=true]{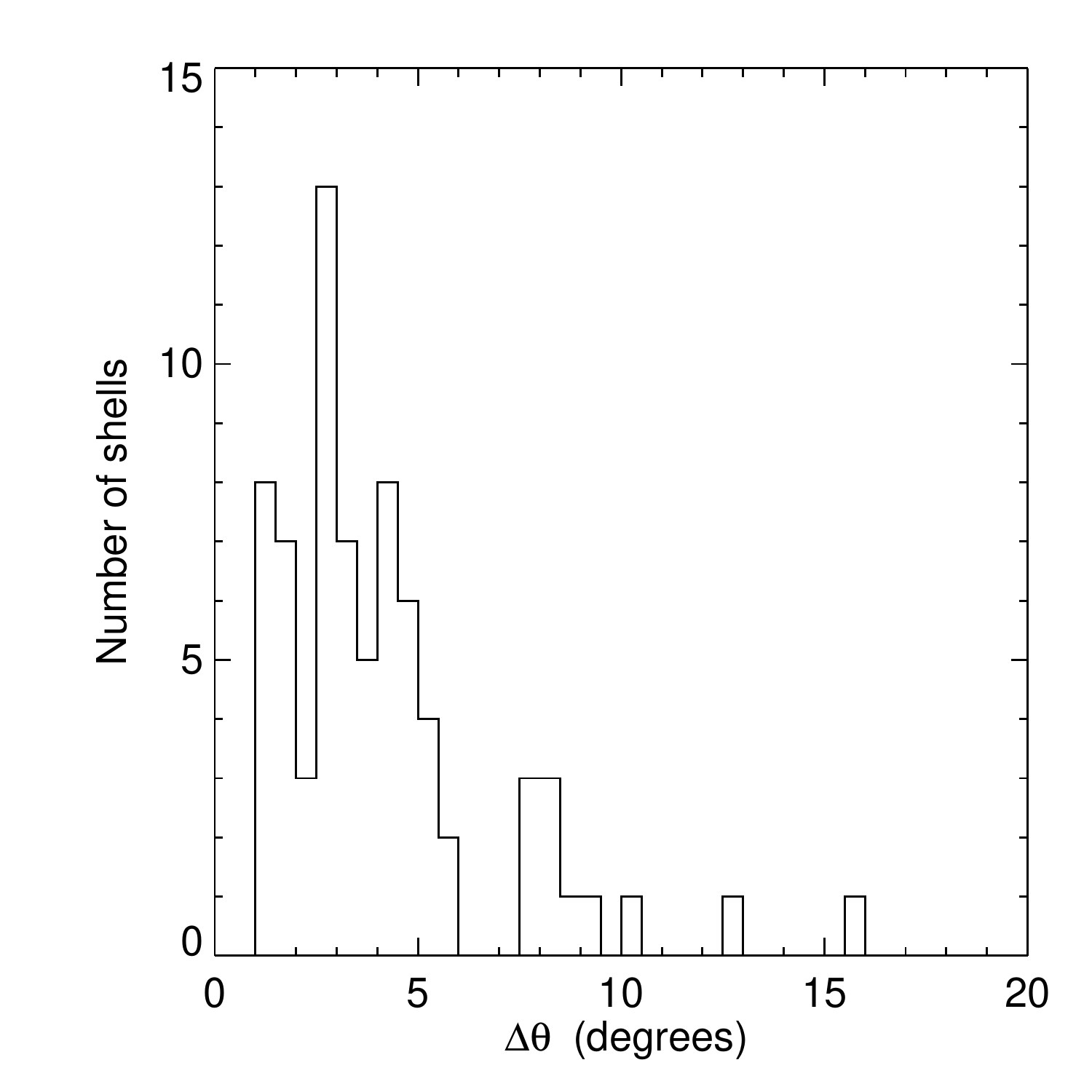}
\else
\includegraphics[width=\columnwidth,keepaspectratio=true]{f5.eps}
\fi
\caption{Distribution of Mean Angular Diameters ($\Delta\theta$) for the 74
shells in our catalog.
\label{fig:MADhist}}
\end{figure}

The shape parameters range from 0.44 to 0.99.  The average $S$ is 0.78, 
while the median is 0.79.  Recall that $S=0$ for a thin
line and $S=1$ for a completely round shell, although non-oval shapes
can make this parameter misleading.  Our catalog contains many
shells with high shape parameters, possibly reflecting the fact
that round features are more likely to be visually identified
as potential shells.  Extremely elongated features (which would have
low $S$) were deemed poor shell candidates, so are not in our 
catalog. 

\begin{figure}[tbp]	
\ifpdf
\includegraphics[width=\columnwidth,keepaspectratio=true]{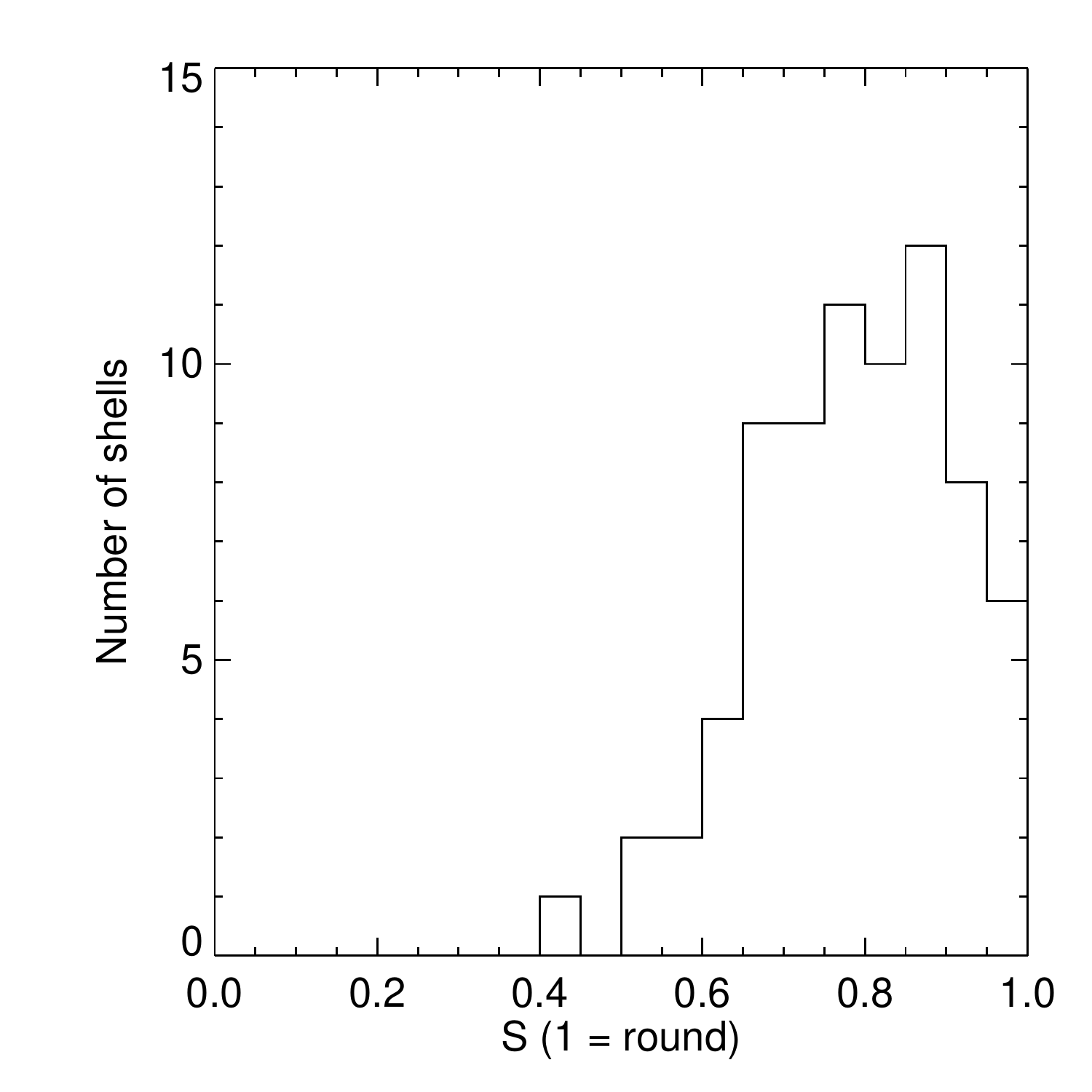}
\else
\includegraphics[width=\columnwidth,keepaspectratio=true]{f6.eps}
\fi
\caption{Distribution of Shape Parameters ($S$) for the 74 shells in our 
catalog.
\label{fig:Shapehist}}
\end{figure}

Distributions of other measured quantities (not shown) were also
examined. The shell wall completeness ($f_{\rm closed}$) is always above
0.5, with a median value of 0.8, reflecting a visual-search bias towards 
regular shell walls. The distribution of  morphological
consistency with velocity ($C$) extends to values as low as 0.1, but
the median is relatively high (0.7).  Without strong additional
indicators, structures with low $C$ values were likely
to be deemed overlapping gas features, rather than shells.
On average, shells in our catalog were maximally
closed over about 60\% of their velocity range, 
with a standard deviation of 17\%.  The distributions
of expansion-related parameters ($C_{\theta}$, $W$, and $PV$) 
all skewed towards lower values, approximately mirroring the $Q_{\rm
exp}$ distribution.

We explored the possibility of correlation between mean angular diameter 
$\Delta\theta$ and shape parameter $S$, as well as whether either 
were correlated with Galactic latitude. None
were observed, with linear correlation coefficient 
$r < 0.25$ in all cases.

\subsubsection{Distances and Physical Sizes}\label{sect:kind}

For those shells with $|b| < 15\arcdeg$, we estimated
a kinematic distance based on $V_{\rm ref}$ and the Galactic rotation
curve of \citet{bb93}.  Velocities of some shells are
inconsistent with Galactic rotation, so no kinematic distance determination
was possible.  In the inner galaxy there are two 
potential solutions, so two distances were 
determined where feasible.  A range of possible distances ($D_{\rm min}$ to
$D_{\rm max}$) was
estimated for each, using $|V_2 - V_1|$ and
assuming the shell was centered on $V_{\rm ref}$.
Although $V_{\rm ref}$ is not always the center of the shell's range of
visibility, it is adequate in light of the
uncertain nature of kinematic distances.  

Peculiar motions complicate kinematic distance determination.  
Possible errors due to non-circular gas motions are not incorporated in
the quoted distance ranges.  \citet{bb93} estimate one-dimensional 
streaming motions in the Galaxy at $\sim$ 12 \kms, 
which adds substantial uncertainty in the derived distances.
For example, \citet{xu06} estimated the peculiar motion of star-forming 
region W3OH at 22 \kms. This caused the kinematic and parallax distances
to differ by about a factor of two (4 kpc vs. 2 kpc).

Of the 33 shells with $b < 15\arcdeg$, kinematic distances were derived 
for 27.  Towards the anti-center ($l \sim$  165$\arcdeg$ to 195$\arcdeg$), 
irregular gas motions may dominate the radial velocity effects of Galactic 
rotation, so results are less reliable.  In Table \ref{table:kin_d},
columns 1-3 contain the Shell identifier and its Galactic coordinates, while  
columns 4-6 contain the kinematic distance $D_{\rm ref}$ corresponding to 
$V_{\rm ref}$ and the range of allowed distances ($D_{\rm min}$ to 
$D_{\rm max}$). Columns 10-12 contain the second set of estimates
for shells in the inner galaxy. Columns 7-9 and
13-15 contain size (diameter) estimates ($S_{\rm ref}$, $S_{\rm min}$, and
$S_{\rm max}$) determined from the corresponding kinematic distances 
and the shell's mean angular diameter.  
Dashes (-) in columns 2-15 indicate 
no solution was found for that value.  Column 16 contains a key to comments; 
most discuss the likelihood
of kinematic distance options based on Galactic spiral structure and/or
inferred shell sizes. Unless otherwise specified, shells are
clearly associated with Galactic plane gas, so kinematic distances are 
relatively reliable.  For convenience during later
discussion, values for shells with single or preferred distance 
estimates are boldfaced in the table.

\subsubsection{Shell Expansion}\label{sect:vexp}

Knowing a shell's expansion velocity is key for understanding its 
evolutionary state.  We therefore derived spectra for
all 18 shells showing moderate signs of expansion ($Q_{\rm exp} \ge 4$).  
We defined both a
spectral extraction region centered on the shell and a constant-$b$ 
background extraction region at $b_{\rm shell}$ outside the shell boundary.
The spectral extraction
region's size was scaled to a fraction of the shell's mean angular diameter
($\Delta\theta$), as were the background region's thickness, minimum
distance and maximum distance from the shell center.  
At every velocity, raw and background spectral values were determined by 
averaging within each region.  We then analyzed the normalized 
(raw $-$ background) spectrum, identifying the local minimum nearest 
$V_{\rm ref}$ and the local maxima on either
side, to estimate the velocities of the shell's approaching
and receding walls.  For noisy spectra, we disregarded spurious
local maxima and estimated the velocity of the appropriate peak(s). 
For each shell, this 
procedure was performed for three different choices of spectral and 
background extraction regions.

We examined the results in detail for every shell.
This included careful evaluation of
the appropriateness of the background region, especially 
where the shell is elongated towards it.
Slight alterations were implemented if
necessary to obtain a physically plausible scenario.
In a few cases we repeated the analysis after slightly adjusting the 
location of the spectral extraction region, based on the shell's 
shape and location variations with velocity.
To evaluate their reasonableness and consistency, 
spectral analysis results were compared against
one another, and against the shell region's visual appearance 
at each velocity using $<$kvis$>$.
This often revealed that gas producing a spectral 
maximum was not unambiguously associated with the shell, and sometimes
that it was clearly unassociated. In some cases, other adjacent 
spectral maxima were clearly identified as the shell wall.
In addition, complex features of varying density in the background 
region obviously introduced spurious features in the 
normalized spectra of certain shells. If possible we adjusted the background 
region to minimize these effects, but often no alternative 
was available. In those cases we based our analysis on the local minima and 
maxima of the raw spectra.

For 7 shells, we identified both front and rear shell walls, and calculated 
$V_{\rm exp} = (V_b - V_f)/2$.  
For a few, the identification of these features is relatively 
unambiguous. The spectra and identified minima/maxima are displayed
in Figure \ref{fig:expvel} for GSH 054-00+003, whose
expansion velocity ($V_{\rm exp} = 9$ \kms) is most certain.
For the rest, unrelated clumps of gas might be
confusing the spectrum, resulting in inaccurate estimates. For these
we also present a lower limit on the expansion velocity based on the best
wall identification and the furthest edge of the velocity range at which 
the structure appeared ``shell-like" ($V_{\rm exp} > (V_b - V_1)/2$ or 
$V_{\rm exp} > (V_2 - V_f)/2$).
For 5 other shells we were able to identify only the front wall or the rear
wall, so could merely derive a lower limit to the expansion velocity.
Inability to identify a front or rear wall for a shell is not unexpected,
both because limb brightening makes them weak relative to the visible walls,
and because high radial velocity dispersion in shell walls would produce
a broad, weak spectral feature.

\begin{figure}[tbp]	
\ifpdf
\includegraphics[width=\columnwidth,keepaspectratio=true]{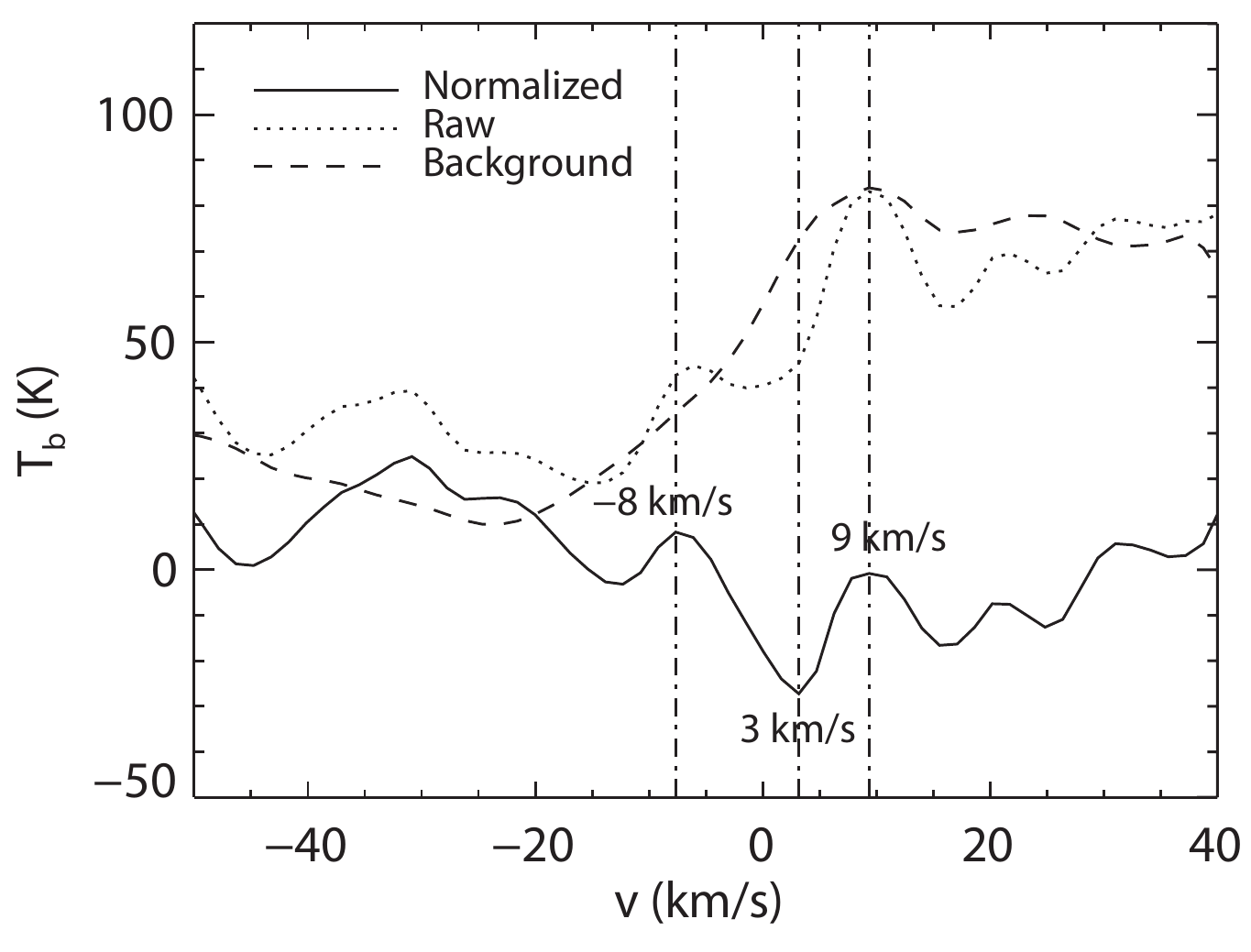}
\else
\includegraphics[width=\columnwidth,keepaspectratio=true]{f7.eps}
\fi
\caption{Raw (dotted) and background-normalized (solid) \HI\ spectra through the center
of GSH 054-00+003. The background (dashed) spectrum is from a constant-$b$
region outside the shell, and was used to normalize the raw spectrum.
The dash-dotted vertical lines mark the velocities of the front and rear walls,
as well as the minimum \HI\ intensity.
\label{fig:expvel}}
\end{figure}

For 6 of the 18 shells we could not significantly constrain the
shell expansion velocity: GSH 048-05+045, GSH 060+01-076,
GSH 110-35-034, GSH 113-54-005, GSH 180-31+020, and GSH 208+32+006.
In these, spectral features potentially denoting shell walls might
be faint and uncertain, confused with other complex gas structures,
in regions/velocities with poor data quality, or a combination of
these factors.  Due to the lack of limb-brightening at the shell center,
sensitivity limitations might prevent us from identifying front or
rear walls, even if the shell is expanding.  In addition, if front or
rear walls are incomplete or contain significant velocity dispersion,
their spectral signature will be difficult to identify. 

Table \ref{table:expvel} displays the results. The first two
columns contain the Shell ID and expansion quality parameter $Q_{\rm
exp}$. Columns 3-5 contain the velocities of the approaching front
and receding back (rear) walls ($V_f$, and $V_b$, respectively), and
the resulting expansion velocity estimate.  Columns 6-8 contain the
shell-like velocity range $V_1$ \& $V_2$, and the lower limit on $V_{\rm
exp}$ described above. The last column contains a key to
comments regarding the process.

The quoted expansion velocities are often quite uncertain. Given the
velocity resolution of the spectra, the best-case scenario is an
uncertainty in both $V_f$ and $V_b$ of $\sim 1.5$ \kms, giving
an uncertainty of at least $\sim 1$ \kms\ for $V_{\rm exp}$. However,
this is overly optimistic due to the image and
spectral complexities described above. For GSH 054-00+003, 
we estimate an expansion velocity uncertainty of 1.5 \kms, with
substantially larger errors for all other shells.  We have 
greater confidence in the quoted lower limits; 
the actual expansion velocity is likely no more than $\sim$ 1 \kms\ below
these, although the possibility remains if we have mis-identified
unrelated gas as part of the shell wall.

\subsection{Other Derived Properties}

Estimating the ages and energies of shells is important for understanding
shell evolution and the changing ISM.  Where we had shell physical size 
and expansion velocity information, we evaluated these quantities as follows. 
Note that we utilized only sizes derived from the
actual kinematic distance estimates, and not the range of possible values.

For shells with estimates of both physical diameter $S_{\rm ref}$ and 
expansion velocity $V_{\rm exp}$, we calculated an upper limit on shell age 
by assuming it has always expanded at its current rate, rather
than slowing over time: $t_1 = S_{\rm ref} / 2 V_{\rm exp}$.  
Where two distance possibilities produced two estimates of shell size,
a second age $t_2$ was also calculated. In cases where we only
estimated a lower limit on $V_{\rm exp}$, we used that 
to estimate the age upper limit(s).

For each shell with a physical size estimate, we used 
$S_{\rm ref}$ to estimate its current thermal energy, assuming
shells are spherical, old and in rough pressure equilibrium with
the ISM.  For a shell of volume $V$, the approximate interstellar gas
pressure $P_{th}/k \sim 3000$ cm$^{-3}$~K \citep{wolfire03} gives a
thermal energy of  $E_{th} = (3/2) n k T V = (3/2) P_{th} V$.  Where we
had two size estimates, we calculated $E_{th}$ for
both.  This energy estimate scales with the assumed pressure
and is very uncertain, as the ISM is far from uniform. In
addition, young shells with larger expansion velocities may 
not yet be in pressure equilibrium.

For shells with an expansion velocity (not just lower
limits), we estimated the current kinetic energy as
$E_K = 0.5 M_{\rm shell} V_{\rm exp}^2$. To evaluate the mass,
we assumed the shell has been expanding into a constant-density
ambient medium, and has swept up all the material in its spherical
volume.  We further assumed that the hydrogen density
is $n$ = 1 atom cm$^{-3}$ and $n_{He} / n = 0.1$, but the results
scale linearly with $n$, so can be easily altered for different
density assumptions.  This energy estimate is necessarily 
crude, given the non-spherical nature of most shells, the 
non-uniformity of the ISM, and the inherent uncertainty
in shell sizes derived from kinematic distances.

For the same shells, we also calculated the initial total shell energy
using Chevalier's formula for a shell of radius $r_{sh}$
now expanding with velocity $V_{\rm exp}$ into an ambient medium of
hydrogen density $n$ \citep{chev74}, again assuming $n$ = 1 atom cm$^{-3}$.  
\begin{displaymath} 
{\rm E_{Ch}} = 5.3 \times 10^{43} \left(\frac{n}{\rm cm^{-3}}\right)^{1.12}
\left(\frac{r_{sh}}{\rm pc}\right)^{3.12} 
\left(\frac{V_{\rm exp}}{\rm km s^{-1}}\right)^{1.4} {\rm ergs} 
\end{displaymath} 
Note that this formula is most reliable for smaller less evolved shells,
but can be an underestimate of the energy of larger shells, especially
those with irregular shapes.

Table \ref{table:age_energy} displays derived physical properties of
these shells, with Shell IDs in the first column. Columns 2 and 6
contain the age upper limits $t_1$ and $t_2$ corresponding to the 
two shell diameter estimates $S_{\rm ref}$ and $S_{\rm ref2}$, 
respectively.  Columns 3 and 7 display the two thermal energy estimates 
${\rm E \times (P / 3000\ cm^{-3} K)}$ corresponding to the two shell sizes, 
and columns 4 and 8 present the corresponding kinetic energy estimates
${\rm E_K \times (1\ cm^{-3} / n)}$. The total energy estimate(s)
${\rm E_{Ch} \times (1\ cm^{-3} / n)}$ are in columns 5 and 9.
Column 10 contains 
codes indicating extra information, such as which values are 
more likely based on kinematic distance preferences described in 
Table \ref{table:kin_d}, or how age upper limits were derived
from expansion velocity (limit) estimates. For convenience during later
discussion, the age and energy values are boldfaced 
for shells with single or preferred distance and size estimates. 

\subsection{Discussion of Shell Properties}

For the majority of shells in Table \ref{table:kin_d}, there
is either only one distance solution, or other considerations such as
Galactic spiral structure led us to prefer one solution. The
exceptions are GSH 040+04+048, GSH 045+14+031, and GSH 049+08+026. We restrict
the remaining discussion to the other 24 shells, for which the preferred 
distances suggest diameters ranging from 18 pc to 1500 pc, with the median 
at 440 pc, and the peak of the distribution at 250 pc.  These shells are 
generally quite large, with only two having diameters $<$ 200 pc, and only 7 
smaller 
than 300 pc.  Biases in our search procedure, and the elimination of 
previously known structures, limit the information present in this distribution.  
However these sizes are similar to those shown in Figure 6
of \citet{ehl05}, and many of the larger sizes were judged plausible
given the apparent shell size relative to the Galactic plane,
as mentioned in the table notes.  Shells this large must be caused by
strong stellar winds and the supernovae of many stars.  For comparison,
the Loop I superbubble, likely generated by the stars of the Sco-Cen
OB association, is $\sim$ 200 pc in diameter \citep{nish99}. Its 
expansion velocity of $\sim 20$ \kms\ suggests it is several million 
years old, although its X-ray intensity suggests an age of $\sim 10^5$ yr.

To evaluate whether the shells were preferentially elongated either
perpendicular or parallel to the Galactic plane, we took the four
locations used to estimate the long and short axes (as described in Section 
\ref{subsec:process}), converted them from 
equatorial to Galactic coordinates, then calculated the angle between the 
long axis and the Galactic plane.
Table \ref{table:PA} contains the resulting semi-major axis ($a$), 
semi-minor axis ($b$), and position angle of the major axis
($\phi = \arctan(\Delta b / \Delta l \cos(b))$) in degrees for all 74 
shells.  The position angles span 0$\arcdeg$ to 180$\arcdeg$, 
showing no preference for alignment parallel or perpendicular to the 
Galactic plane, and we see
no evidence for variations in alignment with angular size $\Delta\theta$.
This conflicts with previous work \citep{ehl05,EP2013,Suad2014} that
suggested most shells are elongated parallel to the plane.  Since there
is very little overlap between our catalog and those (and none with
\citet{ehl05}), our search criteria and process may mean we are looking 
at a slightly different population of interstellar structures.
We also suggest that the trend in orientation found by others might
result from a selection bias, as the low-$|b|$ edge of a
structure could be truncated in visual or automated searches due to the
increasing gas density towards the plane. Finally, many shells are highly
irregular in shape, severely complicating orientation analyses based on
$\Delta b / \Delta l \cos(b)$, elliptical fits, or visual identification
of major \& minor axes.

The expansion velocities and limits in Table \ref{table:expvel} are all
larger than 5 \kms, and mostly greater than 10 \kms. Recall, however,
that expansion velocities were not determined for the majority of
shells in the catalog, which show little or no signs of expansion. 
Those presented are therefore unlikely to be typical of
the overall shell population. We also note that all expansion velocity
measurements are less than 20 \kms, indicating these shells are 
relatively evolved. The corresponding ages presented in Table
\ref{table:age_energy} are generally quite large, over 3.5 Myr in all but
one case (GSH 054-00+003, but the larger distance and age is preferred).
These are upper limits, however, and the actual age is likely less than
about half of the values given.

To put shell energy estimates into context, 
a typical supernova initially has $\sim 10^{51}$ ergs of kinetic energy.
However, heat, interstellar turbulence, and radiation have dissipated
most of this energy for older, evolved shells. By the time a supernova
remnant (SNR) has slowed to 10 \kms, the kinetic energy of a shell is
reduced to less than about 10\% of its initial value, and can be much 
lower if the density of the ambient medium is sufficiently high 
\citep[e.g.][]{chev74,Spitzer78,Thornton98}. For an evolved single-supernova
shell, we therefore expect current energy estimates of order 
$\sim 10^{50}$ ergs.

We have a total of 37 current thermal energy estimates for 27 shells, 
24 of which have either only one energy estimate, or one that is preferred
as noted in Table \ref{table:kin_d}.
Similarly, we have a total of 10 current kinetic energy estimates for 6
shells, 5 with single or preferential values.
In what follows, we consider only these single or preferential energy estimates
(bold-faced in Table \ref{table:age_energy}).
The median thermal energy estimate is 8.6 $\times 10^{50}$ ergs, 
while the median kinetic energy estimate is 1.0 $\times 10^{51}$ ergs.
Nearly all these current shell energies are consistent with an 
evolved shell produced by at least one supernova, often more. 
The few shells with much lower thermal energy values may 
be a result of inaccurate kinematic distances, the crude nature of our 
energy estimates, and/or a different evolutionary history for the shell.
For shells expanding into the Warm Ionized Medium (WIM), the ambient density
could be as low as $n = 0.1$ for some shells, which would decrease the 
calculated kinetic energies by a factor of 10. For this density, the 
kinetic energies of GSH 054-00+003 and GSH 054+01+031 fall below $10^{50}$
ergs; however these shells appear embedded near Galactic plane gas, so
the ambient density is likely higher than this minimum value.

In contrast to the thermal and kinetic energy estimates, 
E$_{Ch}$ estimates the energy required to create each 
structure. 
We evaluated this total energy for 6 shells, 5 of which have single 
or preferential distances/sizes. For these 5, 
the median value of E$_{Ch}$ is 1.3 $\times 10^{52}$ ergs,
with all of the estimates greater than $10^{51}$ ergs.
In general, this suggests that most shells
required the energy input of multiple supernovae for their formation,
however all energy estimates assumed $n = 1$ hydrogen
 atom cm$^{-3}$.  For the smallest likely ambient density in the WIM
($n = 0.1$), the calculated total energies would decrease to 8\% of 
their tabulated values. In this case $E_{Ch}$ for two
shells would fall below $10^{51}$ ergs, to $\sim 3 \times 10^{50}$ ergs. 
However as noted earlier, both these shells lie in or near higher-density 
Galactic plane material. In addition, recall that E$_{Ch}$, based on 
single-supernova models, can be an underestimate for large shells such 
as these.

\subsection{GSH 054-00+003}

We now consider GSH 054-00+003 in more detail, in the context of SNR
modeling. Recall from Tables \ref{table:kin_d}, \ref{table:expvel}
and \ref{table:age_energy} that it has an expansion velocity of 9 \kms,
a radius of 120pc, a current kinetic energy $E_K = 2.1 \times 10^{50}$
ergs, and a total energy E$_{Ch} = 3.8 \times 10^{51}$ ergs, taking the
preferred distance and assuming $n_H = 1$ cm$^{-3}$.  
If the energy of a single SN is 
E$_{SN} = 10^{51}$ ergs, E$_{Ch}$ suggests several supernovae
together produced GSH 054-00+003.  According to the models of Figures 8
and 9 of \citet{Thornton98}, a supernova expanding into material with
this density will grow to $r_{sh} \sim 75$ pc by the time it
has slowed to $V_{\rm exp} \sim 10$ \kms. At this time ($\sim$ 2 Myr)
the SNR will have a kinetic energy $E_{K,mod} \sim 3.8 \times 10^{49}$
ergs (4\% of its initial $E_K$). Thus $\sim 6$ supernovae
would be required to produce the current kinetic energy of GSH 054-00+003,
while a comparison of the observed and model shell volumes would suggest
$\sim 4$ supernovae are required. These are consistent with the estimate
based on E$_{Ch}$.

We chose a hydrogen density of $n$ = 1 cm$^{3}$ because this shell is 
clearly embedded in
relatively dense Galactic plane gas near a spiral arm. In addition,
Thornton \etal's models suggest that for $n \sim 0.1$ cm$^{-3}$
(appropriate for the Warm Ionized Medium), a single SNR would be 
$r_{sh} \sim 140$ pc by the time it has slowed to
$V_{\rm exp} \sim 20$ \kms, thus becoming larger than the observations
for GSH 054-00+003 before slowing to the observed expansion rate. Also,
E$_{Ch}$ for this case is less than $10^{51}$ ergs.  But
if the embedding gas has a density of $n = 10$, 
the
models of \citet{Thornton98} suggest a single-supernova SNR will have a
radius of 30pc and a kinetic energy of $2.5 \times 10^{49}$ ergs when its
expansion has slowed to the observed value ($\sim$ 0.6 Myr), in which case
the observed volume and current kinetic energy (which scales as $n$)
suggest $\sim$ 60-90 supernovae are required to form the shell. The
estimated total shell energy E$_{Ch}$ in this case is $\sim 50$ times
the energy of a single supernova.

Finally, note that the large age estimates are more consistent with 
the absence of enhanced 0.25 keV X-ray emission \citep{snowden97}
from the direction of GSH 054-00+003 (accessed using {\it SkyView};
\citet{mcglynn98}).

\section{Conclusions \& Future Work}

Our visual search of the \SETHI\ database resulted in the 
identification of 74 previously unknown interstellar shells. 
The catalog is uniquely based on high-resolution data that is not limited to
the Galactic plane, and 
is not biased against older non-expanding shells, unlike many
earlier searches. It
is also more sensitive to irregular shells with fragmented walls
than most automated searches.  We presented basic measurements 
(position, reference velocity, angular size, elongation, position angle) 
for all 74 shells, along with kinematic distances, physical sizes,
and expansion velocities where possible.  Shells in the catalog with 
kinematic distances are large, old, and expanding relatively 
slowly if at all. Energy considerations suggest they all formed
by multiple supernovae.  In contrast to findings by others, our 
catalog shells are not preferentially
elongated either parallel to or perpendicular to the Galactic plane.

The \SETHI\ dataset has better angular resolution than the LAB/LDS used
by most previous large-scale searches, and includes high Galactic latitudes,
unlike other high-resolution \HI\ surveys. The GALFA-\HI\ Survey now provides 
higher-quality data with these characteristics, but was not available when this 
labor-intensive project 
began. The GALFA data are currently best at declinations complementary
to those we examined here, so a future search based on
GALFA data will add to the census of Galactic \HI\ shells.

Galactic radial distribution, size distribution, and filling factor of
\HI\ shells are of interest for modeling the Galactic ISM. However,
this catalog is incomplete on its own. Where there is spatial overlap, 
combining it with other large-scale catalogs would require 
careful consideration of the different nature of the underlying data. 
The identification biases and sensitivity limitations are also
drastically different for the varied search techniques. 
In addition, size
and shape measurements based on equatorial coordinates are difficult to 
directly compare with those of catalogs delimited in Galactic coordinates.
To assist with this issue, we provided the estimates of semi-major/minor 
axes and position angles.  If such
an integration were to be carried out, it would also enhance data
on the distribution of shell sizes and their number relative to distance
from the Galactic plane, or Galactic longitude.






\acknowledgments

During this work Dr. Korpela was supported in part by NASA grant NNX09AN69G and
NSF grant AST-0709347. Dr. Sallmen was supported in part by NSF grant AST-0507326, the Research Corporation (Cottrell College Science Award No. CC6476), and
by NASA/Wisconsin Space Grant Consortium's Research Infrastructure Award Program.

The \SETHI\ Survey was funded by the National Science Foundation through grant
AST-0307956 with technical
support provided by the staff of the Arecibo Observatory.  The Arecibo 
Observatory is part of the National Astronomy and Ionosphere Center
which was, during the course of this work, operated by Cornell University and
Univerisities Space Research Association under Cooperative Agreements with the
National Science Foundation.

Kevin Douglas produced the 7.68\degrees$\times$7.68\degrees\ 
\SETHI\ data cubes that were combined and utilized in this analysis.  
La Crosse undergraduate students Lillian Kasel, Tyler Laszczkowski, and high 
school student Daniel Morrison assisted with portions of the shell 
measurements. 

We acknowledge the use of NASA's {\it SkyView}
facility (\url{http://skyview.gsfc.nasa.gov}) located at NASA Goddard Space 
Flight Center.  This research has made use of the SIMBAD database, operated
at CDS, Strasbourg, France \citep{Simbad}.




Facilities: \facility{Arecibo}.

\ifpp{
\tabletypesize{\scriptsize}
\begin{table*}[tb]
\begin{center}
\caption{Catalog of \SETHI\ Shells\label{table:shell_list}}
\begin{tabular}{lcccccccccccc}
\hline
Shell ID & RA & Dec & V$_{\rm ref}$ & $\Delta$RA & $\Delta$Dec & V$_1$ & V$_2$ & $\Delta\theta$ & 
S  & Q & Q$_{\rm exp}$ & Comment \\
 & (hh mm) & ($\arcdeg$) & (\kms) & ($\arcdeg$) & ($\arcdeg$) & (\kms) & (\kms) & ($\arcdeg$) &  &  &  & \\
\hline\hline
GSH 029+34+005  &  16 42  &  12.25  &  5  &  2  &  1.75  &  0  &  12  &  1.8  &  0.81  &  7.9  &  1.7  &    \\
GSH 029+38+005  &  16 28  &  13.75  &  5  &  4  &  2.75  &  2  &  6  &  3.1  &  0.76  &  8.0  &  1.7  &    \\
GSH 030+67-006  &  14 33  &  23.5  &  -6  &  2.75  &  4  &  -9  &  -1  &  3.2  &  0.72  &  7.0  &  2.0  &  \\
GSH 034+20+011  &  17 42  &  9.5  &  11  &  3.5  &  2.5  &  8  &  19  &  2.8  &  0.71  &  6.0  &  2.3  &    \\
GSH 035+36+005  &  16 42  &  16.75  &  5  &  5.75  &  5  &  2  &  11  &  4.8  &  0.85  &  6.2  &  3.0  &    \\
GSH 039+49-017  &  15 52  &  24.25  &  -17  &  9  &  8  &  -23  &  -12  &  8.1  &  0.96  &  7.0  &  2.3  &    \\
GSH 040+04+048  &  18 49  &  8.25  &  48  &  1.75  &  1.75  &  39  &  54  &  1.7  &  0.85  &  7.5  &  5.8  &    \\
GSH 042+21+019  &  17 50  &  17  &  19  &  13.25  &  11.5  &  12  &  25  &  12.8  &  0.70  &  8.4  &  1.7  &  * \\
GSH 044+00-025  &  19 11  &  9.75  &  -25  &  1.25  &  1.5  &  -31  &  -18  &  1.1  &  0.66  &  7.1  &  5.5  &    \\
GSH 044+38+002  &  16 45  &  24.75  &  2  &  2  &  2.5  &  -1  &  5  &  1.8  &  0.78  &  9.5  &  1.0  &    \\
GSH 045+14+031  &  18 23  &  17  &  31  &  2  &  2.75  &  28  &  39  &  2.4  &  0.66  &  6.4  &  4.0  &    \\
GSH 048-05+045  &  19 36  &  11.25  &  45  &  6.75  &  4.5  &  40  &  51  &  4.9  &  0.64  &  6.4  &  4.7  &  * \\
GSH 049+08+026  &  18 51  &  18.5  &  26  &  2.50  &  2.5  &  23  &  29  &  2.5  &  0.96  &  6.2  &  3.0  &  * \\
GSH 052-05+023  &  19 45  &  14  &  23  &  8  &  8.25  &  17  &  39  &  7.5  &  0.85  &  7.5  &  6.3  &  * \\
GSH 052+01+012  &  19 26  &  17.5  &  12  &  6.5  &  5.25  &  11  &  16  &  5.2  &  0.81  &  6.5  &  2.0  &    \\
GSH 052+02-071  &  19 19   &  17.75  &  -71  &  4.75  &  4.25  &  -77  &  -63  &  4.4  &  0.86  &  7.3  &  1.8  &    \\
GSH 052+10-087  &  18 48  &  21.25  &  -87  &  1  &  0.75  &  -91  &  -82  &  1.0  &  0.95  &  8.7  &  6.0  &    \\
GSH 052+20+012  &  18 10  &  25  &  12  &  3.75  &  3  &  9  &  17  &  2.7  &  0.77  &  7.8  &  5.2  &    \\
GSH 054-00+003  &  19 31  &  18.25  &  3  &  1.5  &  1.5  &  -1  &  6  &  1.4  &  0.84  &  7.1  &  5.8  &    \\
GSH 054+01+031  &  19 27  &  19  &  31  &  2  &  1.75  &  28  &  37  &  1.7  &  0.99  &  8.5  &  6.5  &    \\
GSH 055+18-005  &  18 20  &  27.25  &  -5  &  6  &  4  &  -9  &  -3  &  4.5  &  0.82  &  7.6  &  0.7  &    \\
GSH 056-06+033  &  19 57  &  18  &  33  &  3.75  &  3  &  23  &  43  &  3.0  &  0.66  &  6.4  &  3.3  &  * \\
GSH 056+02-074  &  19 28  &  21.75  &  -74  &  1.4  &  1.5  &  -77  &  -62  &  1.4  &  0.70  &  6.9  &  6.8  &  * \\
GSH 057+04+005  &  19 24  &  23.25  &  5  &  6.25  &  6.5  &  2  &  9  &  5.9  &  0.92  &  7.2  &  1.7  &  * \\
GSH 057+12-077  &  18 50  &  26.5  &  -77  &  1.75  &  1.25  &  -80  &  -68  &  1.1  &  0.85  &  8.3  &  5.3  &    \\
GSH 060+01-076  &  19 40   &  24.5  &  -76  &  3.5  &  2.5  &  -80  &  -68  &  2.7  &  0.83  &  7.2  &  4.0  &    \\
GSH 061-01+000  &  19 50.5  &  23.75  &  0  &  1.5  &  1.5  &  -9  &  6  &  1.4  &  0.82  &  7.9  &  3.7  &    \\
GSH 062+00+045  &  19 48  &  25.25  &  45  &  3.5  &  3.5  &  37  &  50  &  2.7  &  0.68  &  6.6  &  2.7  &  * \\
GSH 062+03-102  &  19 35  &  27  &  -102  &  2  &  1.75  &  -107  &  -99  &  1.5  &  0.86  &  7.8  &  3.0  &    \\
GSH 063+00-022  &  19 49  &  26.5  &  -22  &  2.25  &  3  &  -25  &  -18  &  2.3  &  0.67  &  8.4  &  3.7  &  * \\
GSH 064-24+011  &  21 18  &  13.5  &  11  &  3.5  &  4.5  &  3  &  14  &  3.8  &  0.79  &  7.6  &  2.5  &  * \\
GSH 072-30+017  &  21 56  &  15  &  17  &  2.75  &  3.25  &  14  &  25  &  3.1  &  0.83  &  7.4  &  1.3  &    \\
GSH 080-22+002  &  21 52  &  25.5  &  2  &  9.25  &  9.5  &  -5  &  8  &  7.7  &  0.85  &  7.8  &  1.2  &    \\
GSH 109-35-011  &  00 01  &  26.25  &  -11  &  6.75  &  4.5  &  -15  &  -6  &  4.9  &  0.55  &  6.2  &  1.8  &    \\
GSH 110-35-034  &  00 05.5  &  26.5  &  -34  &  3.25  &  $>$ 5  &  -43  &  -28  &  4.0  &  0.75  &  7.7  &  6.7  &    \\
GSH 112-46-008  &  00 19  &  16.25  &  -8  &  3.9  &  5  &  -14  &  -5  &  4.4  &  0.87  &  5.6  &  1.0  &    \\
GSH 113-54-005  &   00 29  &  8.5  &  -5  &  3.25  &  3  &  -9  &  0  &  2.8  &  0.74  &  7.8  &  6.7  &    \\
GSH 116-49-006  &  00 34  &  13.5  &  -6  &  3.25  &  3  &  -9  &  -5  &  2.7  &  0.75  &  7.9  &  0.7  &    \\
GSH 124-52-008  &  00 53  &  10.5  &  -8  &  2.75  &  2.75  &  -14  &  -5  &  2.7  &  0.86  &  6.0  &  2.3  &    \\
GSH 134-43-062  &  01 26  &  19.5  &  -62  &  11.75  &  9  &  -68  &  -51  &  8.4  &  0.74  &  4.5  &  0.3  &  * \\
GSH 139-37+006  &  01 46  &  24  &  6  &  5.25  &  4.5  &  3  &  11  &  4.3  &  0.81  &  8.2  &  2.5  &    \\
GSH 155-32+005  &  02 47  &  23.5  &  5  &  6.25  &  6  &  3  &  9  &  5.0  &  0.90  &  7.5  &  3.0  &  * \\
GSH 156-37-003  &  02 43  &  19  &  -3  &  6.25  &  4.5  &  -6  &  0  &  4.6  &  0.67  &  7.4  &  3.3  &  * \\
GSH 157-27-045  &  03 05  &  26.5  &  -45  &  6  &  3.75  &  -54  &  -31  &  4.4  &  0.75  &  6.7  &  1.3  &  * \\
GSH 170-21+020  &  04 06  &  23.75  &  20  &  5.5  &  4  &  14  &  25  &  4.1  &  0.91  &  8.0  &  3.0  &    \\
GSH 179-24+012  &  04 17  &  15.5  &  12  &  20.75  &  11.75  &  11  &  14  &  15.9  &  0.72  &  9.1  &  3.3  &    \\
GSH 180-31+020  &  04 00  &  10  &  20  &  4.25  &  3.5  &  16  &  25  &  3.8  &  0.98  &  6.3  &  4.5  &  * \\
GSH 182-18+005  &  04 46  &  17.5  &  5  &  4.25  &  3.75  &  3  &  9  &  3.8  &  0.96  &  9.4  &  2.0  &    \\
GSH 183-16-031  &  04 55  &  18  &  -31  &  10.25  &  11  &  -37  &  -20  &  10.3  &  0.92  &  6.4  &  0.7  &    \\
GSH 185-07-009  &  05 31  &  20.5  &  -9  &  4.25  &  5.25  &  -18  &  -3  &  4.2  &  0.65  &  7.0  &  2.0  &    \\
GSH 187-12+012  &  05 18  &  16.5  &  12  &  9.25  &  9.75  &  9  &  19  &  9.1  &  0.93  &  8.1  &  2.0  &    \\
GSH 187+01+020  &  06 05  &  23  &  20  &  3  &  3  &  16  &  26  &  2.9  &  0.81  &  8.3  &  2.7  &    \\
GSH 188+07-079  &  06 31  &  25.5  &  -79  &  3  &  6.25  &  -88  &  -73  &  4.5  &  0.51  &  4.6  &  2.7  &    \\
GSH 190-02+025  &  05 59  &  19.25  &  25  &  2  &  2.25  &  23  &  31  &  2.2  &  0.83  &  7.1  &  0.7  &    \\
GSH 192+06-017  &  06 33  &  21  &  -17  &  3.75  &  4  &  -28  &  -6  &  3.6  &  0.86  &  8.5  &  1.0  &  * \\
GSH 193-01+026  &  06 09  &  17.25  &  26  &  1.75  &  1.25  &  3  &  31  &  1.3  &  0.76  &  7.1  &  5.3  &    \\
GSH 196+10+008  &  06 57  &  19  &  8  &  5.75  &  5  &  6  &  12  &  5.0  &  0.88  &  7.5  &  1.3  &    \\
GSH 197-02+034  &  06 16  &  13  &  34  &  2.75  &  2.75  &  31  &  45  &  2.5  &  0.68  &  6.2  &  0.8  &  * \\
GSH 197+00+002  &  06 23  &  14.5  &  2  &  3  &  2  &  -3  &  5  &  2.7  &  0.78  &  7.2  &  1.0  &    \\
GSH 198+01+034  &  06 25  &  13.75  &  34  &  2.75  &  4.75  &  29  &  39  &  3.2  &  0.44  &  6.3  &  2.8  &  * \\
GSH 198+03-018  &  06 33  &  14.5  &  -18  &  2.5  &  1.25  &  -20  &  -12  &  1.8  &  0.68  &  6.7  &  3.5  &    \\
GSH 200+01-015  &  06 30  &  11.75  &  -15  &  1.75  &  2  &  -20  &  -8  &  1.7  &  0.77  &  7.6  &  1.7  &    \\
GSH 208+32+006  &  08 43  &  18  &  6  &  7.75  &  4.5  &  2  &  17  &  5.5  &  0.67  &  6.8  &  4.0  &    \\
GSH 210+54-003  &  10 12  &  23.25  &  -3  &  3.5  &  3.5  &  -9  &  2  &  2.6  &  0.65  &  7.2  &  3.0  &    \\
GSH 213+28+012  &  08 34  &  12  &  12  &  4.25  &  3.25  &  8  &  20  &  3.7  &  0.77  &  5.9  &  0.0  &    \\
GSH 221+60+000  &  10 45  &  19.25  &  0  &  3.5  &  2.75  &  -3  &  3  &  2.6  &  0.85  &  7.4  &  2.7  &    \\
GSH 225+55-005  &  10 30  &  15.5  &  -5  &  3.25  &  3.25  &  -9  &  -1  &  3.4  &  0.92  &  6.8  &  0.8  &    \\
GSH 228+80-040  &  12 08  &  24  &  -40  &  7.75  &  8.5  &  -48  &  -29  &  7.8  &  0.94  &  7.8  &  5.0  &    \\
GSH 231+55-009  &  10 36  &  12.5  &  -9  &  3.75  &  3  &  -12  &  -6  &  3.0  &  0.71  &  6.8  &  0.0  &    \\
GSH 236+75-008  &  11 51.5  &  20.25  &  -8  &  1.75  &  1.5  &  -12  &  -3  &  1.1  &  0.94  &  8.0  &  1.0  &    \\
GSH 261+74-025  &  12 08  &  15  &  -25  &  11  &  5.25  &  -29  &  -20  &  8.0  &  0.63  &  7.7  &  3.3  &  * \\
GSH 262+73+003  &  12 05  &  13.5  &  3  &  11.5  &  7  &  2  &  8  &  8.6  &  0.56  &  7.3  &  1.0  &    \\
GSH 274+74-006  &  12 19  &  12.5  &  -6  &  10  &  3.5  &  -14  &  -3  &  5.4  &  0.52  &  4.9  &  0.3  &    \\
GSH 294+76+000  &  12 43  &  13.25  &  0  &  4.5  &  5.5  &  -1  &  8  &  4.3  &  0.72  &  5.4  &  3.2  &    \\
\end{tabular}
\end{center}
\end{table*}

\tabletypesize{\scriptsize}
\begin{table*}[tb]
\begin{center}
\caption{ Physical Characteristics \label{table:kin_d}}
\begin{tabular}{lccccccccccccccc}
\hline
Shell ID & $l$ & $b$ & D$_{\rm ref}$ & 
D$_{\rm min}$ & D$_{\rm max}$ & S$_{\rm ref}$ & S$_{\rm min}$ & S$_{\rm max}$ &
D$_{\rm ref2}$ & D$_{\rm min2}$ & D$_{\rm max2}$ & S$_{\rm ref2}$ & S$_{\rm min2}$ & S$_{\rm max2}$ & 
Comment \\
 & ($\arcdeg$) & ($\arcdeg$) & (kpc) & (kpc) & (kpc) & (pc) & (pc) & (pc) & 
(kpc) & (kpc) & (kpc) & (pc) & (pc) & (pc)  & \\
\hline\hline
GSH 040+04+048  &  40.01  &  4.29  &  3.3  &  2.7  &  3.9  &  98  &  80  &  120  &  9.8  &  9.2  &  10  &   290  &  270  &  300 &    \\
GSH 044+00-025  &  43.83  &  0.14  &  {\bf 15}  &  14  &  15  &  {\bf 290}  &  270  &  290  &  -  &  -  &  -  &   -  &  -  &  - &    \\
GSH 045+14+031  &  45.19  &  13.81  &  2.3  &  1.8  &  2.8  &  96  &  75  &  120  &  10  &  9.6  &  11  &   420  &  400  &  460 &  \tablenotemark{c} \\
GSH 048-05+045  &  48.05  &  -4.57  &  3.5  &  2.9  &  4.2  &  300  &  250  &  360  &  {\bf 7.9}  &  7.2  &  8.5  &   {\bf 680}  &  620  &  730 &  \tablenotemark{a}\tablenotemark{d} \\
GSH 049+08+026  &  49.47  &  8.41  &  1.9  &  1.6  &  2.3  &  83  &  70  &  100  &  9.2  &  8.9  &  9.6  &   400  &  390  &  420 &    \\
GSH 052-05+023  &  51.54  &  -5.14  &  1.7  &  0.75  &  2.8  &  220  &  98  &  370  &  {\bf 8.9}  &  7.8  &  9.9  &   {\bf 1200}  &  1000  &  1300 &  \tablenotemark{a}\tablenotemark{d} \\
GSH 052+01+012  &  52.38  &  0.55  &  0.81  &  0.55  &  1.1  &  74  &  50  &  100  &  {\bf 9.6}  &  9.3  &  9.8  &   {\bf 870}  &  840  &  890 &  \tablenotemark{b} \\
GSH 052+02-071  &  51.82  &  2.14  &  {\bf 19}  &  17  &  20  &  {\bf 1500}  &  1300  &  1500  &  -  &  -  &  -  &   -  &  -  &  - &    \\
GSH 052+10-087  &  51.67  &  10.24  &  {\bf 22}  &  21  &  24  &  {\bf 380}  &  370  &  420  &  -  &  -  &  -  &   -  &  -  &  - &  \tablenotemark{c}\tablenotemark{f} \\
GSH 054-00+003  &  53.61  &  -0.14  &  0.05  &  -  &  0.43  &  1.2  &  -  &  11  &  {\bf 10}  &  9.7  &  10  &   {\bf 240}  &  240  &  240 &  \tablenotemark{g} \\
GSH 054+01+031  &  53.81  &  1.05  &  2.5  &  2  &  3.2  &  74  &  59  &  95  &  {\bf 7.5}  &  6.9  &  8.1  &   {\bf 220}  &  200  &  240 &  \tablenotemark{a} \\
GSH 056-06+033  &  56.47  &  -5.62  &  3.1  &  1.8  &  -  &  160  &  94  &  -  &  {\bf 6.4}  &  -  &  7.7  &   {\bf 340}  &  -  &  400 &  \tablenotemark{h} \\
GSH 056+02-074  &  56.34  &  2.16  &  {\bf 18}  &  16  &  19  &  {\bf 440}  &  390  &  460  &  -  &  -  &  -  &   -  &  -  &  - & \tablenotemark{d} \\
GSH 057+04+005  &  57.23  &  3.68  &  0.18  &  -  &  0.59  &  19  &  -  &  61  &  {\bf 9.1}  &  8.6  &  9.5  &   {\bf 940}  &  890  &  980 &  \tablenotemark{a} \\
GSH 057+12-077  &  56.72  &  12.03  &  {\bf 19}  &  18  &  20  &  {\bf 360}  &  350  &  380  &  -  &  -  &  -  &   -  &  -  &  - &  \tablenotemark{c}\tablenotemark{d} \\
GSH 060+01-076  &  60.09  &  1.08  &  {\bf 17}  &  16  &  18  &  {\bf 800}  &  750  &  850  &  -  &  -  &  -  &   -  &  -  &  - &  \tablenotemark{d} \\
GSH 061-01+000  &  60.64  &  -1.37  &  -  &  -  &  0.55  &  -  &  -  &  13  &  {\bf 8.6}  &  7.8  &  9.4  &   {\bf 210}  &  190  &  230 &  \tablenotemark{b} \\
GSH 062+03-102  &  61.72  &  3.28  &  {\bf 21}  &  20  &  22  &  {\bf 550}  &  520  &  580  &  -  &  -  &  -  &   -  &  -  &  - &  \tablenotemark{d}\tablenotemark{i} \\
GSH 063+00-022  &  62.84  &  0.32  &  {\bf 10}  &  9.7  &  10  &  {\bf 400}  &  390  &  400  &  -  &  -  &  -  &   -  &  -  &  - &  \tablenotemark{d} \\
GSH 187-12+012  &  187.1  &  -12.05  &  {\bf 8.4}  &  3.5  &  20  &  {\bf 1300}  &  560  &  3200  & ... & ... & ... & ...  & ... & ... &    \\
GSH 187+01+020  &  187.32  &  0.76  &  {\bf 24}  &  9.3  &  -  &  {\bf 1200}  &  470  &  -  & ... & ... & ... & ...  & ... & ... &  \tablenotemark{e} \\
GSH 190-02+025  &  189.89  &  -2.31  &  {\bf 18}  &  11  &  33  &  {\bf 690}  &  420  &  1300  & ... & ... & ... & ...  & ... & ... &  \tablenotemark{e} \\
GSH 193-01+026  &  192.8  &  -1.22  &  {\bf 11}  &  3  &  -  &  {\bf 250}  &  68  &  -  & ... & ... & ... & ...  & ... & ...&  \tablenotemark{e} \\
GSH 196+10+008  &  196.47  &  9.72  &  {\bf 1.5}  &  0.79  &  2.4  &  {\bf 130}  &  69  &  210  & ... & ... & ... & ...  & ... &    &  \tablenotemark{c}\tablenotemark{d} \\
GSH 197-02+034  &  197.34  &  -1.78  &  {\bf 10}  &  6.4  &  17  &  {\bf 440}  &  280  &  740  & ... & ... & ... & ...  & ... & ... &  \tablenotemark{e} \\
GSH 197+00+002  &  196.82  &  0.43  &  {\bf 0.38}  &  -  &  1.2  &  {\bf 18}  &  -  &  57  & ... & ... & ... & ...  & ... & ...&    \\
GSH 198+01+034  &  197.71  &  0.51  &  {\bf 9.8}  &  7.2  &  14  &  {\bf 550}  &  400  &  780  & ... & ... & ... & ...  & ... & ... &  \tablenotemark{e} \\
\end{tabular}
\end{center}
\begin{description}
\item[a] Greater distance more likely based on Galactic spiral structure in direction of shell
\item[b]Greater distance somewhat preferred based on Galactic spiral structure and/or shell size
\item[c] Not clear if shell is associated with local or Galactic gas. Kinematic distances could well be inaccurate.
\item[d] Inferred shell size plausible in context of extent of Galactic plane
\item[e] Seems associated with spiral structure but kinematic distances
are too large for spiral structure in this direction (actually at 2.5 kpc not 10
kpc). May be affected dramatically by streaming motions. Kinematic distances
could well be inaccurate.
\item[f] Fairly high above Galactic plane given the kinematic distance.
Kinematic distances could well be inaccurate.
\item[g] Appears to be in the nearer of 2 spiral arms. Likely the larger distance because a shell $<$ 10pc in size is unlikely if embedded in a spiral arm.
\item[h] Allowed distance ranges for the two solutions overlap. Likely nearer the upper edge of range based on spiral arm locations in this direction.
\item[i] Appears to be part of 3$^{\rm rd}$ spiral arm along line of sight (recently discovered to extend in this direction).
\end{description}
\end{table*}

\tabletypesize{\footnotesize}
\begin{table*}[tb]
\begin{center}
\caption{ Expansion Velocities\label{table:expvel}}
\begin{tabular}{lcccccccc}
\hline
Shell ID & $Q_{\rm exp}$ & 
$V_f$ & $V_b$ & $V_{\rm exp}$ &
$V_1$ & $V_2$ & LL on $V_{\rm exp}$ &
Comment \\
 &  &
(\kms) & (\kms) & (\kms) & 
(\kms) & (\kms) & (\kms) &  \\
\hline\hline
GSH 040+04+048  &  5.8 	&  33 	& ...  	&  ... 	&  39	&  54	&  11	& \tablenotemark{d}	\\
GSH 044+00-025	&  5.5  & -43 	& -12 	& 16 	& -31	& -18	&  13  	& \tablenotemark{c}\tablenotemark{d} \\
GSH 045+14+031	&  4.0	&  17 	&  40   & 12   	&  28	&  39	&   6	& \tablenotemark{a}\tablenotemark{b}\tablenotemark{f} \\
GSH 052-05+023	&  6.3	&  10 	&  50 	& 20 	&  17	&  39	&  15	& \tablenotemark{c}\tablenotemark{d}\tablenotemark{e} \\
GSH 052+10-087	&  6.0	&  ...	& -77	& ...	& -91	& -82	&   7	& \tablenotemark{b} \\
GSH 052+20+012	&  5.2	&   0	&  20	& 10	&   9	&  17	&   9	& \tablenotemark{a}\tablenotemark{c}\tablenotemark{d} \\
GSH 054-00+003	&  5.8	& -8	&   9	&  9	&  -1	&   6   &  ...	& 	\\
GSH 054+01+031	&  6.5	&  23 	&  42	& 10	&  28	&  37	&   7	& \tablenotemark{a}\tablenotemark{b}\tablenotemark{c}\tablenotemark{e} \\
GSH 056+02-074	&  6.8	& -80	& -57	& 12	& -77	& -62	&   9 	& \tablenotemark{c}\tablenotemark{d}	\\
GSH 057+12-077	&  5.3	& ... 	& -53	& ...	& -80	& -68	&  14	& \tablenotemark{b}\tablenotemark{f}\tablenotemark{g} \\
GSH 193-01+026	&  5.3	&  -5 	& ...	& ...	&   3	&  31	&  18	& \tablenotemark{d}\tablenotemark{h}	\\
GSH 228+80-040	&  5.0	& -56 	& ... 	& ...	& -48	& -29	&  14  	& \tablenotemark{d}\tablenotemark{e}	\\
\end{tabular}
\end{center}
\begin{description}
\item[a]Front wall estimate may be possibly unrelated gas, or suffer from data quality issues. 
\item[b]Estimated $V_{\rm exp}$ lower limit uses $V_1$ and $V_b$.
\item[c]Rear wall estimate may be possibly unrelated gas, or suffer from data quality issues. 
\item[d]Estimated $V_{\rm exp}$ lower limit uses $V_f$ and $V_2$.
\item[e]Based on raw spectrum.
\item[f]Expansion velocity might be underestimated because
at least one maximum in spectrum is broad and/or peaks at a velocity 
very close to last shell-like velocity.
\item[g]Clear maximum on one side, but no clear minimum in spectrum.
\item[h]Spectrum complex, and constantly changing histogram limits required to
follow its morphology vs. velocity.
\end{description}

\end{table*}

\tabletypesize{\scriptsize}
\begin{table*}
\begin{center}
\caption{ Derived Shell Properties \label{table:age_energy}}
\begin{tabular}{lcccccccccc}
\hline
Shell ID & 
$t_1$ & ${\rm E_1 P^{-1}_{3000 cm^{-3} K}}$ & ${\rm E_{K1}  n^{-1}_{cm^{-3}}}$ & ${\rm E_{Ch1}  n^{-1.12}_{cm^{-3}}}$ & 
$t_2$ & ${\rm E_2 P^{-1}_{3000 cm^{-3} K}}$ & ${\rm E_{K2}  n^{-1}_{cm^{-3}}}$ & ${\rm E_{Ch2}  n^{-1.12}_{cm^{-3}}}$ & 
Comment \\
 & 
(Myr) & (erg) & (erg) & (erg) &
(Myr) & (erg) & (erg) & (erg) & \\
\hline\hline
GSH 040+04+048 &     4.3 &  8.7$\times 10^{48}$  &    ...   &    ...   &     13. &  2.3$\times 10^{50}$  &    ...   &    ...   &  \tablenotemark{a} \\
GSH 044+00-025 &     {\bf 8.6} &  ${\bf 2.1 \times 10^{50}}$  &  ${\bf 1.0 \times 10^{51}}$  &  ${\bf 1.3 \times 10^{52}}$  &   ...   &    ...   &    ...   &    ...   &  \\
GSH 045+14+031 &     3.9 &  8.5$\times 10^{48}$  &  2.3$\times 10^{49}$  &  3.0$\times 10^{50}$  &     17. &  7.1$\times 10^{50}$  &  1.9$\times 10^{51}$  &  3.0$\times 10^{52}$  &\tablenotemark{b}   \\
GSH 048-05+045 &   ...   &  2.5$\times 10^{50}$  &    ...   &    ...   &   ...   &  ${\bf 3.0 \times 10^{51}}$  &    ...   &    ...   &  \tablenotemark{c} \\
GSH 049+08+026 &   ...   &  5.9$\times 10^{48}$  &    ...   &    ...   &   ...   &  6.3$\times 10^{50}$  &    ...   &    ...   &  \\
GSH 052-05+023 &     5.5 &  1.1$\times 10^{50}$  &  8.1$\times 10^{50}$  &  8.7$\times 10^{51}$  &     {\bf 29.} &  ${\bf 1.5 \times 10^{52}}$  &  ${\bf 1.1 \times 10^{53}}$  &  ${\bf 1.5 \times 10^{54}}$  & \tablenotemark{c}  \\
GSH 052+01+012 &   ...   &  3.9$\times 10^{48}$  &    ...   &    ...   &   ...   &  ${\bf 6.3\times 10^{51}}$  &    ...   &    ...   & \tablenotemark{c}  \\
GSH 052+02-071 &   ...   &  ${\bf 2.8\times 10^{52}}$  &    ...   &    ...   &   ...   &    ...   &    ...   &    ...   &  \\
GSH 052+10-087 &     {\bf 27.} &  ${\bf 5.8\times 10^{50}}$  &    ...   &    ...   &   ...   &    ...   &    ...   &    ...   & \tablenotemark{a}\tablenotemark{b} \\
GSH 054-00+003 &   0.054 &  9.6$\times 10^{42}$  &  1.5$\times 10^{43}$  &  1.3$\times 10^{44}$  &     {\bf 13.} &  ${\bf 1.4\times 10^{50}}$  &  ${\bf 2.1\times 10^{50}}$  &  ${\bf 3.8\times 10^{51}}$  & \tablenotemark{c}  \\
GSH 054+01+031 &     3.7 &  4.0$\times 10^{48}$  &  7.6$\times 10^{48}$  &  1.1$\times 10^{50}$  &     {\bf 11.} &  ${\bf 1.1\times 10^{50}}$  &  ${\bf 2.0\times 10^{50}}$  &  ${\bf 3.2\times 10^{51}}$  & \tablenotemark{c}  \\
GSH 056-06+033 &   ...   &  3.9$\times 10^{49}$  &    ...   &    ...   &   ...   &  ${\bf 3.6\times 10^{50}}$  &    ...   &    ...   & \tablenotemark{c}  \\
GSH 056+02-074 &     {\bf 18.} &  ${\bf 7.8\times 10^{50}}$  &  ${\bf 2.1\times 10^{51}}$  &  ${\bf 3.3\times 10^{52}}$  &   ...   &    ...   &    ...   &    ...   &  \\
GSH 057+04+005 &   ...   &  6.6$\times 10^{46}$  &    ...   &    ...   &   ...   &  ${\bf 7.7\times 10^{51}}$  &    ...   &    ...   & \tablenotemark{c}  \\
GSH 057+12-077 &     {\bf 13.} &  ${\bf 4.5\times 10^{50}}$  &    ...   &    ...   &   ...   &    ...   &    ...   &    ...   & \tablenotemark{a}\tablenotemark{b}   \\
GSH 060+01-076 &   ...   &  ${\bf 4.8\times 10^{51}}$  &    ...   &    ...   &   ...   &    ...   &    ...   &    ...   &  \\
GSH 061-01+000 &   ...   &    ...   &    ...   &    ...   &   ...   &  ${\bf 8.9\times 10^{49}}$  &    ...   &    ...   & \tablenotemark{c}\tablenotemark{d}   \\
GSH 062+03-102 &   ...   &  ${\bf 1.7\times 10^{51}}$  &    ...   &    ...   &   ...   &    ...   &    ...   &    ...   &  \\
GSH 063+00-022 &   ...   &  ${\bf 6.3\times 10^{50}}$  &    ...   &    ...   &   ...   &    ...   &    ...   &    ...   &  \\
GSH 187-12+012 &   ...   &  ${\bf 2.3\times 10^{52}}$  &    ...   &    ...   &   ...   &    ...   &    ...   &    ...   &  \\
GSH 187+01+020 &   ...   &  ${\bf 1.7\times 10^{52}}$  &    ...   &    ...   &   ...   &    ...   &    ...   &    ...   & \tablenotemark{b}  \\
GSH 190-02+025 &   ...   &  ${\bf 3.1\times 10^{51}}$  &    ...   &    ...   &   ...   &    ...   &    ...   &    ...   & \tablenotemark{b}  \\
GSH 193-01+026 &     {\bf 6.8} &  ${\bf 1.5\times 10^{50}}$  &    ...   &    ...   &   ...   &    ...   &    ...   &    ...   & \tablenotemark{a}\tablenotemark{b}  \\
GSH 196+10+008 &   ...   &  ${\bf 2.2\times 10^{49}}$  &    ...   &    ...   &   ...   &    ...   &    ...   &    ...   & \tablenotemark{b}  \\
GSH 197-02+034 &   ...   &  ${\bf 8.6\times 10^{50}}$  &    ...   &    ...   &   ...   &    ...   &    ...   &    ...   & \tablenotemark{b}  \\
GSH 197+00+002 &   ...   &  ${\bf 5.6\times 10^{46}}$  &    ...   &    ...   &   ...   &    ...   &    ...   &    ...   &  \\
GSH 198+01+034 &   ...   &  ${\bf 1.6\times 10^{51}}$  &    ...   &    ...   &   ...   &    ...   &    ...   &    ...   & \tablenotemark{b}  \\
\end{tabular}
\end{center}
\begin{description}
\item[a]Expansion velocity lower limit used to derive age limits.
\item[b]Treat results with caution, as kinematic distances may be 
inaccurate.
\item[c]Greater age limits and/or energies preferred, based on 
kinematic distance preference noted earlier.
\item[d]Smaller kinematic distance option only gave an upper limit
on distance.
\end{description}

\end{table*}

\begin{table*}[tb]
\begin{center}
\caption{Shell Axes and Position Angles \label{table:PA}}
\begin{tabular}{lccc}
\hline
Shell ID & $a$ & $b$ & $\phi$ \\ 
 & ($^o$) & ($^o$) & ($^o$) \\
\hline\hline
GSH 029+34+005 &     1.1 &     0.7 &  139 \\
GSH 029+38+005 &     1.9 &     1.2 &  134 \\
GSH 030+67-006 &     2.0 &     1.1 &  173 \\
GSH 034+20+011 &     1.8 &     1.0 &   89 \\
GSH 035+36+005 &     2.7 &     2.0 &  137 \\
GSH 039+49-017 &     4.2 &     3.9 &  125 \\
GSH 040+04+048 &     1.0 &     0.7 &  138 \\
GSH 042+21+019 &     8.3 &     4.5 &  155 \\
GSH 044+00-025 &     0.7 &     0.4 &  174 \\
GSH 044+38+002 &     1.1 &     0.7 &    2 \\
GSH 045+14+031 &     1.6 &     0.8 &  174 \\
GSH 048-05+045 &     3.3 &     1.6 &  105 \\
GSH 049+08+026 &     1.3 &     1.2 &   71 \\
GSH 052+01+012 &     3.1 &     2.1 &  104 \\
GSH 052+02-071 &     2.5 &     1.9 &   81 \\
GSH 052+10-087 &     0.5 &     0.5 &  139 \\
GSH 052+20+012 &     1.7 &     1.0 &  144 \\
GSH 052-05+023 &     4.3 &     3.2 &    7 \\
GSH 054+01+031 &     0.9 &     0.8 &   79 \\
GSH 054-00+003 &     0.8 &     0.6 &    5 \\
GSH 055+18-005 &     2.7 &     1.8 &  125 \\
GSH 056+02-074 &     0.9 &     0.5 &  170 \\
GSH 056-06+033 &     2.0 &     1.0 &   82 \\
GSH 057+04+005 &     3.2 &     2.7 &   40 \\
GSH 057+12-077 &     0.7 &     0.5 &  132 \\
GSH 060+01-076 &     1.5 &     1.1 &  125 \\
GSH 061-01+000 &     0.8 &     0.6 &  173 \\
GSH 062+00+045 &     1.8 &     0.9 &   56 \\
GSH 062+03-102 &     0.9 &     0.7 &  172 \\
GSH 063+00-022 &     1.5 &     0.8 &    6 \\
GSH 064-24+011 &     2.3 &     1.5 &   11 \\
GSH 072-30+017 &     1.8 &     1.3 &   58 \\
GSH 080-22+002 &     4.4 &     3.2 &   69 \\
GSH 109-35-011 &     3.5 &     1.3 &  131 \\
GSH 110-35-034 &     2.5 &     1.5 &   84 \\
GSH 112-46-008 &     2.5 &     1.9 &   69 \\
GSH 113-54-005 &     1.8 &     1.0 &  100 \\
GSH 116-49-006 &     1.6 &     1.0 &  174 \\
GSH 124-52-008 &     1.5 &     1.2 &   92 \\
GSH 134-43-062 &     5.3 &     3.1 &   17 \\
GSH 139-37+006 &     2.5 &     1.7 &   43 \\
GSH 155-32+005 &     2.8 &     2.3 &   13 \\
GSH 156-37-003 &     3.1 &     1.5 &   51 \\
GSH 157-27-045 &     2.7 &     1.6 &   60 \\
GSH 170-21+020 &     2.2 &     1.9 &   44 \\
GSH 179-24+012 &    10.2 &     5.7 &   50 \\
GSH 180-31+020 &     1.9 &     1.9 &   48 \\
GSH 182-18+005 &     2.0 &     1.8 &  144 \\
GSH 183-16-031 &     5.6 &     4.7 &  149 \\
GSH 185-07-009 &     2.8 &     1.4 &  175 \\
GSH 187+01+020 &     1.7 &     1.2 &  180 \\
GSH 187-12+012 &     4.9 &     4.2 &  144 \\
GSH 188+07-079 &     3.4 &     1.1 &  148 \\
GSH 190-02+025 &     1.3 &     0.9 &  170 \\
GSH 192+06-017 &     2.1 &     1.6 &  167 \\
GSH 193-01+026 &     0.8 &     0.5 &   61 \\
GSH 196+10+008 &     2.8 &     2.2 &   87 \\
GSH 197+00+002 &     1.6 &     1.0 &   42 \\
GSH 197-02+034 &     1.6 &     0.8 &   15 \\
GSH 198+01+034 &     2.5 &     0.7 &  173 \\
GSH 198+03-018 &     1.2 &     0.6 &   63 \\
GSH 200+01-015 &     1.0 &     0.6 &  166 \\
GSH 208+32+006 &     3.7 &     1.9 &   58 \\
GSH 210+54-003 &     1.7 &     0.8 &  120 \\
GSH 213+28+012 &     2.3 &     1.4 &   40 \\
GSH 221+60+000 &     1.5 &     1.1 &   95 \\
GSH 225+55-005 &     1.8 &     1.5 &   74 \\
GSH 228+80-040 &     4.2 &     3.7 &    3 \\
GSH 231+55-009 &     1.9 &     1.1 &   50 \\
GSH 236+75-008 &     0.6 &     0.5 &  155 \\
GSH 261+74-025 &     5.6 &     2.6 &   41 \\
GSH 262+73+003 &     6.3 &     2.4 &   52 \\
GSH 274+74-006 &     4.0 &     1.4 &   29 \\
GSH 294+76+000 &     2.8 &     1.6 &  126 \\
\hline
\end{tabular}
\end{center}
\end{table*}

}




\end{document}